\documentclass[aps,pra,twocolumn,superscriptaddress,showpacs]{revtex4}

\usepackage{graphicx}
\usepackage{dcolumn}

\newcommand{\rdr}{R_{(7)}}
\newcommand{\rdei}{R_{(3)}}
\newcommand{\rdeii}{R_{(4)}}

\begin{document}

\preprint{To be submitted to Phys. Rev. A}

\title{Dissociative recombination and low-energy inelastic
  electron collisions\\of the helium dimer ion}

\author{H. B. Pedersen}
\thanks{ To whom correspondence should be addressed}
\email[]{henrik.pedersen@mpi-hd.mpg.de}
\author{H. Buhr}

\author{S. Altevogt}
\author{V. Andrianarijaona}
\author{ H. Kreckel}
\author{L. Lammich}
\affiliation{Max-Planck-Institut f\"{u}r Kernphysik, D-69117 Heidelberg, Germany}

\author{N. de Ruette}
\author{E. M. Staicu-Casagrande}
\affiliation{D\'{e}partement de Physique, Universit\'{e} Catholique de Louvain, B-1348, 
Louvain-la-Neuve, Belgium}

\author{D. Schwalm}
\affiliation{Max-Planck-Institut f\"{u}r Kernphysik, D-69117 Heidelberg, Germany}
\author{D. Strasser}
\affiliation{Department of Particle Physics, Weizmann Institute of
Science, Rehovot, 76100, Israel}

\author{X. Urbain}
\affiliation{D\'{e}partement de Physique, Universit\'{e} Catholique de Louvain, B-1348, 
Louvain-la-Neuve, Belgium}

\author{D. Zajfman} 
\affiliation{Max-Planck-Institut f\"{u}r Kernphysik, D-69117 Heidelberg, Germany}
\affiliation{Department of Particle Physics, Weizmann Institute of
Science, Rehovot, 76100, Israel}
\author{A. Wolf}
\affiliation{Max-Planck-Institut f\"{u}r Kernphysik, D-69117 Heidelberg, Germany}
\date{\today}

\begin{abstract}
  
  The dissociative recombination (DR) of $^3$He$^4$He$^+$ has been
  investigated at the heavy-ion Test Storage Ring (TSR) in Heidelberg
  by observing neutral products from electron-ion collisions in a
  merged beams configuration at relative energies from near-zero
  (thermal electron energy about 10 meV) up to 40 eV.  After storage
  and electron cooling for 35 s, an effective DR rate coefficient at
  near-zero energy of $3\times10^{-9}$ cm$^{3}$\,s$^{-1}$ is found.
  The temporal evolution of the neutral product rates and fragment
  imaging spectra reveals that the populations of vibrational levels
  in the stored ion beam are non-thermal with fractions of 
  $\sim$0.1--1\% in excited levels up to at least $v=4$, having a
  significant effect on the observed DR signals. With a
  pump-probe-type technique using DR fragment imaging while switching
  the properties of the electron beam, the vibrational excitation of
  the ions is found to originate mostly from ion collisions with the
  residual gas.  Also, the temporal evolution of the DR signals
  suggests that a strong electron induced rotational cooling occurs in
  the vibrational ground state, reaching a rotational temperature near
  or below 300 K.  From the absolute rate coefficient and the shape of
  the fragment imaging spectrum observed under stationary conditions,
  the DR rate coefficient from the vibrational ground state is
  determined; converted to a thermal electron gas at 300 K it amounts
  to $(3.3 \pm 0.9) \times 10^{-10}$ cm$^{3}$\,s$^{-1}$.  The
  corresponding branching ratios from $v=0$ to the atomic final states
  are found to be
  $(3.7 \pm 1.2)$\% for $1s2s\,{^3}S$, %
  $(37.4 \pm 4.0)$\% for $1s2s\,{^1}S$, %
  $(58.6 \pm 5.2)$\% for $1s2p\,{^3}P$, and %
  $(2.9 \pm 3.0)$\% for $1s2p\,{^1}P$. %
  A DR rate coefficient in the range of $2\times10^{-7}$
  cm$^{3}$\,s$^{-1}$ or above is inferred for vibrational levels $v=3$
  and higher.  
  As a function of the collision energy, the measured DR
  rate coefficient displays a structure around 0.2 eV.  At higher
  energies, it has one smooth peak around 7.3 eV and a highly
  structured appearance at 15--40 eV.
  The small size of the observed
  effective DR rate coefficient at near-zero energy indicates that the
  electron induced rotational cooling is due to inelastic electron-ion
  collisions and not due to selective depletion of rotational levels
  by DR.
\end{abstract}

\pacs{}

\maketitle

\section{Introduction}
\label{sec:intro}


Low-energy collisions of diatomic, positive molecular ions with free
electrons are important processes in cold dilute media, such as the
interstellar medium, planetary atmospheres, and laboratory discharges
where they are among the reactions controlling the degree of
ionization and the chemical composition.  Usually the dominant
neutralization process is the dissociative recombination (DR)
\cite{bates1994} between an incident molecular ion $AB^+$ and a free
electron $e^-$
\begin{equation}
AB^+(v,J) + e^-(E) \rightarrow A(n)+B(n').
\label{eq:general_dr}
\end{equation}
where $v$ and $J$ denote the rovibrational quantum numbers of the
molecular ion, while $E$ and $n, n'$ are the electron energy and the
atomic final-state quantum numbers, respectively.  Beside
recombination, dissociative reactions with charged products exist,
which are represented by
\begin{equation}
AB^+(v,J) + e^-(E) \rightarrow \left\{
\begin{array}{l}A^+(n)+B(n')\\A(n)+B^+(n') \end{array}
\right\} +  e^-(E')
\label{eq:general_de}
\end{equation}
and termed dissociative excitation (DE).  Finally, non-dissociative
inelastic interactions occur, in particular through collisions of the
type
\begin{equation}
AB^+(v,J) + e^-(E) \rightarrow  AB^+(v',J')
+ e^-(E'),
\label{eq:general_ex}
\end{equation}
which describe the electron-impact excitation or de-excitation of
nuclear motion in the molecular ion.

Beyond the relevance of these collision processes in cold dilute media, the
understanding of their underlying quantum dynamical mechanisms
is of fundamental interest.
In particular, non-adiabatic dynamics manifestly violating the 
Born-Oppenheimer approximation often plays an important role, 
which makes low-energy electron collisions of small molecular ions 
a benchmark case for understanding non-adiabatic molecular
interactions involving electronic continuum states.

For DR,  
the attractive interaction in the initial state and the exothermic,
non-radiative nature of the process mostly lead to large cross
sections at low collision energies, often reaching the order of
$10^{-14}$ cm$^2$ for an incident energy of $E\sim0.03$ eV.  For 
many systems, including the experimentally and theoretically well
studied cases of H$_2^+$ and its isotopomers \cite{larsson1997} as
well as heavier systems such as O$_2^+$ \cite{kella1997,guberman1997},
NO$^+$ \cite{vejby1998,sun1990}, and others, a slow incident electron
can be resonantly captured by purely electronic interactions into
doubly excited repulsive states of the neutral system, which enables
the `direct' mechanism \cite{bates1950,bates1994} as an efficient
pathway for DR.  Some systems exist where suitable electronic
potentials for the `direct' process are absent for the low-lying
ro-vibrational states; however, in many such cases, including the
astrophysically important molecular ions HeH$^+$ and H$_3^+$
\cite{larsson1997}, non-adiabatic mechanisms are nevertheless
considered to cause DR with similar cross sections as in the `direct'
process \cite{bates1994,kokoouline2003}.

\begin{figure}[htbp]
\centering
\includegraphics[width=3.5in]{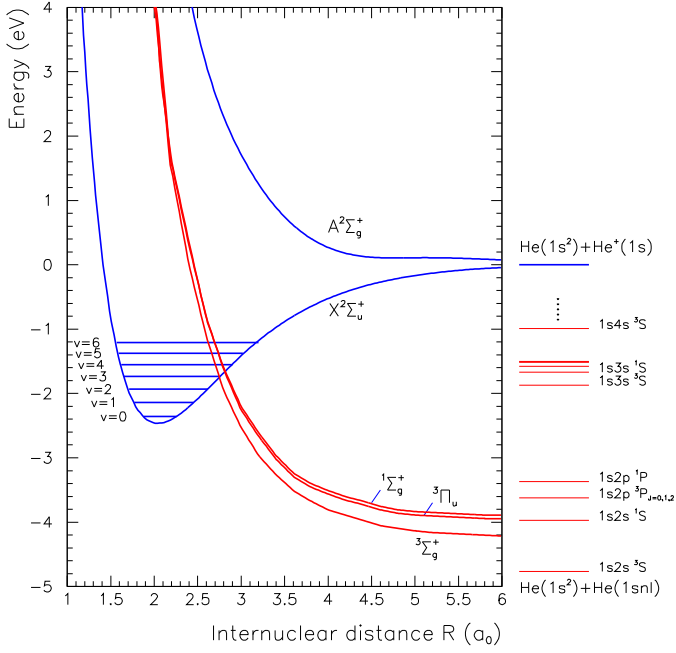}
\caption{
  Potential energy curves for the $X{^2}\Sigma_u^+$ electronic ground
  state of He$_2^+$ \cite{cencek1995} and its first dissociative state
  ($A{^2}\Sigma_g^+$) \cite{ackermann1991} together with the three
  lowest dissociating curves of He$_2$ \cite{cohen1976,carata1999}.
  The energetic positions of the first seven vibrational levels of
  $^3$He$^4$He$^+$ are shown as horizontal lines in the
  $X{^2}\Sigma_u^+$ potential.  Asymptotically, the ${{^3}\Sigma_g^+}$
  state correlates to He$(1s^2\,{^1}S)$ + He$(1s2s\, {^3}S)$, the
  ${{^1}\Sigma_g^+}$ state to He$(1s^2\, {^1}S)$ + He$(1s2s \,{^1}S)$,
  and the ${{^3}\Pi_u}$ state to He$(1s^2\, {^1}S)$ + He$(1s2p\,
  {^3}P)$.  }
\label{he_curves}
\end{figure}

A system falling outside this picture is the helium dimer ion,
He$_2^+$.  From early theoretical work \cite{mulliken1964} up to the
most recent calculation \cite{carata1999}, extremely low cross
sections (of order $10^{-18}$ cm$^2$ for $E\sim0.03$ eV) have been
predicted for the DR of He$_2^+$ in low-lying initial vibrational
states.  Electronic potential curves driving the `direct' DR process
at low electron energies are not accessible from the lower vibrational
states, as shown in Fig.\ \ref{he_curves}.  The inclusion of
rovibrational interactions in theoretical calculations
\cite{carata1999} causes a rich structure of narrow electron capture
resonances, but still yields a small effect in the energy-averaged
rate coefficient; these calculations did not include non-adiabatic
coupling to neutral states below the ion curve as a significant
increase of the cross section due to such coupling was not expected.
For excited vibrational states ($v\gtrsim3$), much larger DR cross
sections ($\sim${}$10^{-14}$ cm$^2$ at $E\sim0.03$ eV as typical for
other species) are predicted \cite{carata1999}.  The low-energy DR of
helium dimer ions should therefore be characterized by extreme
variations of the cross sections among the low vibrational states.

The helium dimer ion is of importance in models of the early universe
\cite{stancil1998} and in laboratory plasmas.  Particularly, He$_2^+$
is the dominant ion in the fundamental helium plasmas and in the
helium afterglow at pressures above $\sim$5 mbar \cite{deloche1976}
and at room temperature (while atomic ions dominate the low-pressure
He afterglows or at higher temperature).  In the helium plasma,
He$_{2}^{+}$ is created in either three-body \cite{phelps1952} or
binary collisions \cite{deloche1976,hornbeck1951} giving ions with a
significant rovibrational excitation \cite{deloche1976,urbain1999a}.
The helium afterglow is characterized by a strong recombination of
electrons and molecular ions and the simultaneous occurrence of
excited (triplet) helium atoms \cite{cheret1972}.  The DR of He$_2^+$
ions was suggested early on \cite{bates1950} to explain the observed
recombination \cite{biondi1949}.  However, later experiments and
modeling suggested collisional-radiative recombination (He$_2^+$ +
2$e^{-}$ $\rightarrow$ He$_2{}^*$ + $e^{-}$) as the dominent process,
with the DR rate coefficient being negligible at an upper limit of
$<${}$5\times10^{-10}$ cm$^3$/s \cite{deloche1976}.  DR from higher
vibrational states (certainly present in the studied high-pressure
afterglows) was also argued to be negligible, while other studies of
the He afterglow and plasma \cite{ivanov1983,ivanov1989} showed a
significant role of DR for vibrationally excited ions in producing
excited He atoms with $n=3,4$.  Thus, the actual role of DR in
high-pressure He afterglows remains unclear.

The structure of the He$_2^+$ ion is well studied theoretically
\cite{cencek1995,ackermann1991} and several experimental studies were
peformed \cite{maas1976,flamme1980,yu1987,coman1999,zande2000,%
  hardy2000}.  In particular, the precise data from vibrational
spectroscopy \cite{yu1987} and the high-level calculations on the
electronic potential energy surface for He$_2^+$ make it possible to
calculate accurate nuclear wavefunctions, reliable vibrational and
rotational radiative lifetimes, and accurate kinetic energy releases
for the DR process, all of which are essential to interpret the
experimental results presented in this paper.

Regarding the low-energy DR process, already Mulliken
\cite{mulliken1964}, on the basis of qualitative arguments on the
electronic structure of He$_2$, suggested that the rate from the
vibrational ground states of He$_2^+$ should be very small, while
still significant for higher vibrational levels.  Calculated diabatic
potential energy curves for dissociative states by Cohen
\cite{cohen1976} and Guberman \cite{guberman1983} confirmed this
hypothesis, since no curve crossing exists close to the ground
vibrational state of the ion and the dissociative curves of the
neutral, thus suppressing the `direct' route of DR.  In a recent paper
Carata {\it et al.}\ \cite{carata1999} presented new Multichannel
Quantum Defect Theory (MQDT) calculations on the low energy DR of
He$_2^+$ addressing several aspects of importance for an experimental
study.  For the three isotopomers, the DR from the vibrational states
$v=0$--4 was calculated (including also rotational excitation)
considering the lowest three neutral dissociative curves of the
symmetries $^3\Sigma_g^+$, $^1\Sigma_g^+$, and $^3\Pi_u$ with the
diabatic asymptotic limits of He($1s^2\,^1S$) + He($1s2s\,^3S$),
He($1s^2\,^1S$) + He($1s2s\,^1S$), and He($1s^2\,^1S$) +
He($1s2p\,^3P$), respectively (see Fig.\ \ref{he_curves}).  At low
relative energies (0--0.1 eV) and for $v=0$--2, the $^3\Sigma_g^+$
curve was found to provide the dominant dissociation route, with a
very small cross section ranging from $\sim${}$10^{-18}$--$10^{-16}$
cm$^2$ for $v=0$, leading to a rate coefficient of only $6.1\times
10^{-11}$ cm$^3$/s at 300 K.  Oppositely, for $v=3$ and 4 the
preferred dissociation route was found to be $^3\Pi_u$ with an average
cross section about four orders of magnitude larger than for $v=0$.
Also the rotational excitation was found to influence the DR cross section,
the result increasing by about a factor of 2 in going from $J=0$ to
$J=9$ for $^3$He$^4$He$^+$($v=0$).  In the work of Carata {\it et
  al.}\ \cite{carata1999} curve crossing in the asymptotic region was
not explicitly considered and therefore the branching ratios into the
final channels were not predicted, however, it was argued that
dissociation should preferentially proceed on the dominant
dissociative curve for a given vibrational level ($^3\Sigma_g^+$ for
$v=0$--2).

\begin{figure}[htbp]
\centering
\includegraphics[width=3.5in]{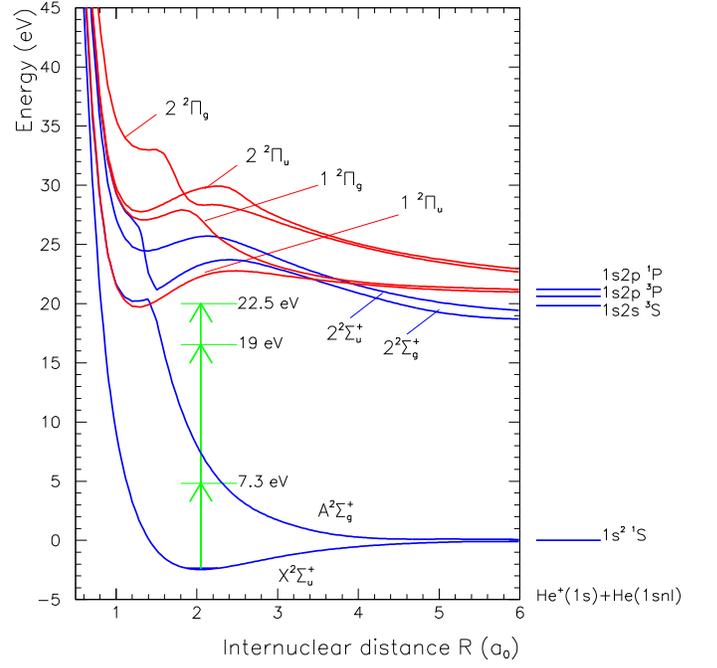}
\caption{
  The lowest doublet potential energy curves of the He$_2^+$ system as
  calculated by Ackermann and Hogreve \cite{ackermann1991}.  The
  vertical arrows indicate the energetic positions above the
  vibrational ground state of $^3$He$^4$He$^+$ where major structures
  in the DR rate coefficient have been observed in the present
  measurement.}
\label{he_curves-high}
\end{figure}

Regarding DR at higher electron energies, the available detailed
calculations for excited electronic states of He$_2^+$
\cite{ackermann1991} indicate at which energies and internuclear
distances doubly excited neutral states, suitable for resonant
electron capture and subsequent dissociation, could be expected.  
An overview of the predicted higher-lying potential curves 
of He$_2^+$ are given in Fig.\ \ref{he_curves-high}.  
Rydberg-like doubly excited neutral states can occur in energy 
regions extending down by $\sim$3 eV from the calculated excited 
ionic potential curves.  Considering
the Franck-Condon zone of He$_2^+$($v=0$), peaks in the DR rate due to
resonant electron capture are expected and in fact observed in the
data presented here at electron energies of $\sim$6--10 eV and further
up at $\gtrsim$20 eV.

In the present experiment, DR and other related electron-induced
reactions are studied for the asymmetric isotopomer $^3$He$^4$He$^+$
of the helium dimer ion, using a stored ion beam and a merged cold
electron beam at the heavy-ion Test Storage Ring (TSR) in Heidelberg
\cite{habs1989}.  The extreme variation of the DR cross section with
the vibrational quantum number imposes large demands on the
understanding of the ion beam evolution during the experiment.  
In particular, the predicted difference in the average DR cross section
of four orders of magnitude \cite{carata1999} in going from $v=0$ to
$v=3$ makes the experimental DR techniques sensitive to
a population of vibrationally excited states of $\lesssim${}$10^{-4}$ 
in the ion beam.

This sensitivity
was already noted in the only earlier merged beams experiment on the
DR of He$_2^+$, performed by Urbain {\it et al.}\ \cite{urbain1999b}
at the ASTRID storage ring using $^3$He$^4$He$^+$ and
$^4$He$^4$He$^+$.  The results for the two species differed strongly
from each other, as expected from their different infrared emission
properties in case of a strong initial vibrational excitation.  For
$^4$He$^4$He$^+$, a DR cross section monotonously decreasing with the
collision energy was observed between 0 and 15 eV with no time
dependence over $\sim$14 s of storage time, while the $^3$He$^4$He$^+$
DR cross section showed a peak at around 6 eV and a strong decrease
with time in the low energy region.  Thus, a high sensitivity of the
reaction rate to the internal excitation, with a small cross section
for lower vibrational states, was demonstrated.

In order to extract quantitative information on the DR process and
related low-energy electron collisions of He$_2^+$, a major part of
the present experimental work is devoted to clarifying the internal
excitation dynamics of an electron cooled $^3$He$^4$He$^+$ ion beam
in a storage ring.  On the one hand, the Coulomb explosion imaging
technique (Sec.\ \ref{sec:exp_CEI}) is used to monitor the vibrational
population of the stored ion beam while, on the other hand, time dependent
measurements of  
DR rate coefficients (Sec.\ 
\ref{sec:dr-rate}) and DR fragment imaging distributions  
(Sec.\ \ref{sec:dr_imaging}) are employed as diagnostic tools to
understand the evolution of the distribution of rovibrational states
in the stored ion beam.

Measurements over long ion storage times (up to 70 s) reveal a
dependence of the low-energy DR rate on the rotational excitation of
the $^3$He$^4$He$^+$ beam and indicate internal cooling of the stored
ion beam by rotationally inelastic electron collisions (Sec.\ 
\ref{sec:total-DR-rate}).

Time dependent observations of structures in the DR fragment imaging
distributions (Sec.\ \ref{sec:initial-to-final}) and pump-probe-type
experiments (Sec.\ \ref{sec:pump-probe}) serve to identify the
vibrational levels contributing to the measured signal; in addition,
they are used to measure the variation of electron-impact vibrational
excitation rates of $^3$He$^4$He$^+$ in an electron energy range of
2--9 eV.

After identifying these contributions, the branching ratios for the
final atomic states
(Sec.\ \ref{initial_final}) and the absolute low-energy DR rate
coefficient (Sec.\ \ref{sec:absrate}) 
are extracted for $^3$He$^4$He$^+$($v=0$) ions.  Energy dependent
absolute cross sections are obtained for low-
energy DR
(Sec.\ \ref{sec:low-e-structure}) 
and for high-energy DR and DE (Sec.\ \ref{sec:high-e-structure},
\ref{sec:res_de_exc}).
Finally, the results on DR are considered in the light of previous
experimental and theoretical data (Sec.\ \ref{sec:discussion_dr}) and the
cross sections of vibrationally and rotationally inelastic collisions,
as implied by the present data, are discussed (Sec.\ 
\ref{sec:discussion_rovib}).

Using the same arrangement and similar procedures, measurements have
also been performed on $^4$He$^4$He$^+$; these results will be
presented in a forthcoming publication.

\section{Experimental background}
\label{sec:experimentalconditions}

\subsection{Ion beam and detector setup}
\label{sec:exp_overview}
\begin{figure*}[hp]
\centering
\includegraphics[width=2.58in,angle=270]{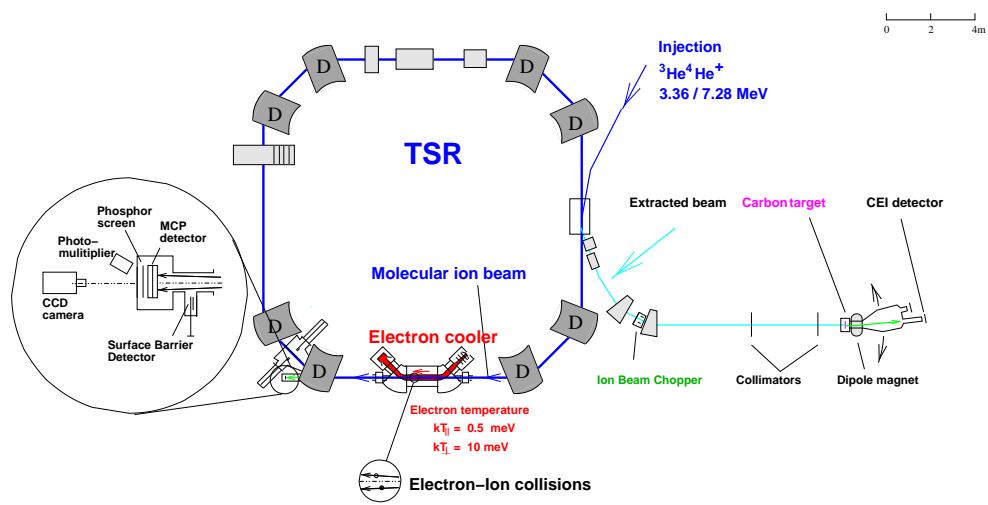}
\caption{
(1.5 columns wide)
Schematic drawing of the experimental setup around the TSR.}
\label{experiment}
\end{figure*}

The experiment was conducted at the Test Storage Ring (TSR)
\cite{habs1989} at the Max-Planck-Institut f\"{u}r Kernphysik in
Heidelberg.  A schematic drawing of the experimental setup is
displayed in Fig.\ \ref{experiment}.  The basic experimental
procedures have been described in detail earlier regarding merged
electron and ion beam experiments in general \cite{kilgus1992}, DR of
molecular ions \cite{amitay1996,alkhalili2003}, and Coulomb explosion
imaging \cite{wester1998,amitay1999}.

For the present experiment, helium dimer ions were produced in a
standard duoplasmatron ion source \cite{green1974}.  Ion source
geometries both with and without an expansion cup following the anode
were tested.  Without an expansion cup the source delivered 300--500
nA of $^3$He$^4$He$^+$ for total pressures of 0.4--1.1 mbar in the
filament chamber and an arc current of 0.33 down to 0.08 mA.  The
currents of atomic $^3$He$^+$ and $^4$He$^+$ were kept about equal at
20--30 $\mu$A.  With an expansion cup in the ion source, the absolute
yields of $^3$He$^4$He$^+$ were considerably lower and the source
showed less stable operation.  During the measurements reported here
we used the source geometry without expansion cup, keeping the
pressure in the cathode chamber at 0.6 mbar and the arc current close
to 0.10 mA.  Weak amounts of N$^{2+}$ were observed in the beam with
the Coulomb explosion technique (Sec.\ \ref{sec:exp_CEI}).  In all the
experiments performed in this work small traces of N$^{2+}$ in the
beam have no influence on the results.

The ions were accelerated to kinetic energies $E_i$ of either 7.28 MeV
or 3.36 MeV using an rf accelerator \cite{grieser1993,vonhahn1993},
transported to the storage ring by means of several magnetic steering
and focusing elements, injected into the storage ring and stored in
the field of its bending and focusing magnets.  Pulses of a few nA
were injected for about 150 $\mu$s and the ions were stored for times
up to 70 s with a mean lifetime (1/$e$) of 9.8 s. The beam loss is
mainly determined by ion collisions with species of the residual gas
which is composed mostly ($\ge 90$\%) of H$_2$.  The average pressure
in the storage ring was 5--$10\times 10^{-11}$ mbar during the present
measurements.

The TSR electron cooler \cite{steck1990,pastuszka1990} was used to
apply phase space cooling to the stored He$_2^+$ ion beam
\cite{poth1990} and to perform the electron-ion recombination
experiments \cite{kilgus1992,amitay1996}.  The electron beam is guided
by a longitudinal magnetic field ($\sim$0.04 T) and has a diameter of
29.5 mm in the region where the electron and ion beam overlap
collinearly, surrounded by a 1.5 m long solenoid magnet.  In two
toroid magnets adjacent to the ``central'' region of the electron
cooler, the electron beam is bent in and out of the overlap region.
Neutral products created by interactions of the He$_2^+$ ions in the
straight section of the storage ring containing the electron cooler
are separated from the circulating ion beam at the downstream
storage-ring magnet (see Fig.\ \ref{experiment}) and detected behind
this magnet as described in detail below.

The relative velocity of electrons and ions in the central part of the
electron cooler is controlled by the electron acceleration voltage
\cite{kilgus1992}.  The difference between the average longitudinal
ion and electron velocities defines the detuning velocity $v_d$ and
the corresponding detuning energy $E_d=(m/2)v_d{}^2$ (where $m$
denotes the electron mass, which here in good approximation can be
used to replace the reduced mass).  The collision velocities ${\bf v}$
are distributed according to an anisotropic Maxwellian
\cite{kilgus1992} characterized by the electron temperatures $T_\perp$
and $T_\parallel$ perpendicular and parallel to the beam direction,
respectively.  For the present experiment $kT_{\perp} = 10$ meV and
$kT_{\parallel} = 0.5$ meV.

The average beam velocities can be exactly matched ($E_d=0$) chosing a
laboratory electron energy $E_e^c=(m/M)E_i$ ($M$ denoting the ion
mass).  At this setting called the ``cooling'' energy, efficient
translational (phase-space) cooling of the stored ion beam takes place
through Coulomb scattering \cite{poth1990} between the ions and the
colder and continuously renewed electrons.  The time needed for
completion of the phase space cooling depends on the initial ion
temperature, the electron intensity and the detailed injection
conditions; for the present measurements the phase space cooling was
completed after 6 s, at the latest.  For energy dependent studies of
electron-ion interactions, the laboratory electron energy $E_e$ can be
detuned from $E_e^c$, yielding finite detuning energies $E_d$ for the
relative motion between electrons and ions as further discussed in
Sec.\ \ref{sec:dr-rate}.

At velocity matching ($E_d=0$), electrons and ions interact with
typical collision energies $E$ in the 10-meV range, corresponding to
the transverse electron temperature ($kT_{\perp}$).  At an ion energy
of $E_i=7.28$ MeV, the cooling energy of the electron beam was
$E_e^c=571$ eV and electron densities $n_e$ of $1.2\times10^7$
cm$^{-3}$ and $5.5\times10^6$ cm$^{-3}$ were used, as specified below.
At $E_i=3.36$ MeV, the cooling energy was $E_e^c=263$ eV and the
electron density was $5.5\times10^6$ cm$^{-3}$.  

Applying electron
energies $E_e>E_e^c$, relative energies $E_d$ up to $\sim$50 eV were
adjusted, resulting in electron collision energies with a
FWHM spread \cite{kilgus1992} of, e.g., $\sim$0.2 eV at $E_d=10$ eV.
Variations of $E_d$ are connected with slight but significant
variations of electron density $n_e(E_d)$.

Two types of detectors are used behind the storage-ring dipole magnet
(Fig.\ \ref{experiment}) to count and analyze the neutral products
created by interactions of the $^3$He$^4$He$^+$ ions with the merged
electron beam or with residual gas molecules.  An energy-sensitive
surface barrier detector with a size of $40\times60$ mm$^2$, centered
to the beam axis, served to count and discriminate events with neutral
products.  With essentially unit efficiency, each event creates an
output pulse of this detector whose height is proportional to the
total kinetic energy of all neutral products.  As all neutral products
propagate with the ion beam velocity, the pulse height thus represents
the total mass of neutral fragments, corresponding to 3, 4, or 7
atomic mass units (amu) if a single $^3$He atom, a single $^4$He atom,
or two ($^3$He + $^4$He) atoms arrive in the neutral product channel.
The interpretation of the corresponding, mass-discriminated neutral
count rates $\rdei$, $\rdeii$, and $\rdr$ in terms of rate
coefficients for electron and residual gas induced processes will be
discussed below.

Instead of the surface barrier detector, also a multi-particle imaging
detector \cite{amitay1996} could be used in the detector chamber.  
From an 80-mm diameter multichannel plate, equipped with a phosphor screen, 
the transverse coordinates of all fragments arriving within a coincidence
time window of a few microseconds, is read out with an event-triggered 
camera system.  This system is used
for DR fragment imaging as described in Sec.\ \ref{sec:dr_imaging}.

\subsection{Event rates for electron-induced processes}
\label{sec:rate_coeff}

The elementary rate coefficient $\alpha_X(E_d)$ of any
electron-induced process $X$, as represented by Eqs.\ 
(\ref{eq:general_dr})--(\ref{eq:general_ex}), is related to the cross
section $\sigma_X(E)$ of this process for a given collision energy
$E=(m/2){\bf v}^2$ through the average over the electron velocity
distribution according to
\begin{equation}
\alpha_X(E_d)=\int |{\bf v}| \sigma_X(E) f(v_d,{\bf v})  d^3v ,
\label{eq:alpha_general}
\end{equation}
where $f(v_d,{\bf v})$ is the electron velocity distribution
\cite{kilgus1992} at a given detuning velocity $v_d=(2E_d/m)^{1/2}$.
In the present work it will be important for some of the
electron-induced processes to consider the dependence of their rate on
the rovibrational state $v,J$ of the stored molecular ions.  In this
context, the ensemble of $N_i$ stored ions will be described by normalized
populations $p_{vJ}(t)$ in the various ro-vibrational levels, which in
general can depend on the storage time (time since injection) $t$.
Introducing state dependent elementary rate coefficients
$\alpha_X(E_d)\equiv \alpha_X^{vJ}(E_d)$ the observed rate
coefficients will be the ro-vibrationally averaged quantities
\begin{equation}
\tilde{\alpha}_X(E_d) = \sum_{vJ}p_{vJ}(t)\alpha_X^{vJ}(E_d).
\label{eq:alp_mean_general}
\end{equation}
Using these rate coefficients, the rate $R_X$ of an electron-induced
process $X$ occuring in the central overlap region of the electron
cooler can be written as
\begin{equation}
R_X = \eta N_i n_e(E_d) \tilde{\alpha}_{X}(E_d).
\label{eq:rx_central}
\end{equation}
Here, $\eta=L/C$, where $L$ is the length of the central overlap
region (1.5 m in the present experiment) and $C$ the storage ring
circumference (55.4 m, yielding $\eta=0.027$).  The expression given 
in Eq.\ (\ref{eq:rx_central}) is not complete, as additional electron-induced
reactions occur in the toroid regions where the electron beam, over a
part of its bent section, still overlaps the ion beam.  In contrast to
the central region, where a single controlled value of $E_d$ occurs,
the detuning energy in the toroid regions rises fast as a consequence
of the rapidly increasing angle between the two beams
\cite{lampert1996,amitay1996}.  In the present measurements, the beams
still overlap on a length of $\sim$16 cm in each of the bends adjacent
to the central section, and the excess detuning energy in these
regions rises by up to $\sim$10 eV.  The rate $R_X$ will hence include
an additional toroid contribution $\tilde{\alpha}^{\rm tor}_{\rm
  X}(E_d)$, so that Eq.\ (\ref{eq:rx_central}) should be replaced by
\begin{equation}
R_X = \eta N_i n_e(E_d) \left[ \tilde{\alpha}_{X}(E_d) 
+ \tilde{\alpha}^{\rm tor}_{X}(E_d) \right].
\label{eq:rx_toroid}
\end{equation}
with \cite{lampert1996,alkhalili2003}
\begin{equation}
\tilde{\alpha}^{\rm tor}_{X}(E_d) = \frac{2}{L}
\int_{x_{\rm min}}^{x_{\rm max}} \tilde{\alpha}_{X}
\biglb( \tilde{E}_d(x;E_d) \bigrb) dx.
\label{eq:alpha_tor}
\end{equation}
Here, $\tilde{E}_d(x;E_d)$ denotes the shifted detuning energy as a
function of the longitudinal position $x$ in the toroid region,
obtained from the known beam geometry.  From an energy dependent rate
measurement $R_X(E_d)$ covering a suitably wide range of detuning
energies $E_d$ it is possible to obtain the toroid contribution
$\tilde{\alpha}^{\rm tor}_{\rm X}(E_d)$ and hence the corrected rate
coefficient $\tilde{\alpha}_{\rm X}(E_d)$ through a robust iterative
procedure \cite{lampert1996,alkhalili2003}.

\subsection{Stored ion beam evolution}
\label{sec:Ion_beam_evolution}

Understanding the temporal evolution of the rovibrational level
distribution in the stored and electron cooled ion beam is essential
for the interpretation of the experimental results presented. Hence,
in this section we outline and model the main processes that lead to
beam loss and influence the rovibrational populations.  In particular,
the rovibrational states are coupled to the $\sim$300 K blackbody
radiation field in the storage ring.  Moreover, the stored ions
experience collisions with electrons in the electron cooler and with
species in the residual gas leading to loss, excitation and
de-excitation of vibrational and rotational motion.  To serve in the
discussion of experimental data, expressions for the time dependence
of the stored ion number, the neutral fragment count rates, and the
rovibrational populations will be derived.

\subsubsection{Beam loss and neutral fragment count rates}
\label{sec:beam_loss}

The relevant processes leading to ion loss from the circulating beam
are inelastic collisions with residual gas molecules and with
electrons in the electron cooler.  Beam loss through elastic
scattering processes can be neglected.  In collisions with a residual
gas particle ($R$), the dominant loss arises through the DE-type
reactions
\begin{equation}
^3{\rm He}^4{\rm He}^+ + R \rightarrow \left\{
  \begin{array}
  {@{\,}l@{\,}c@{\,\,}l@{}}^3{\rm He}^+&+&^4{\rm He}\\^3{\rm
    He}&+&^4{\rm He}^+ 
  \end{array}
\right\} + R'.
\label{eq:de_res}
\end{equation}
For each of these processes we define a rate constant $k_{\rm DE}^{g}$, which 
describes the reaction rate per stored ion and depends on the 
reaction cross section and the gas density.  
Residual gas
collsions can also lead to dissociative charge exchange (DC)
\begin{equation}
^3{\rm He}^4{\rm He}^+ + R \, \rightarrow\, ^3{\rm He}\,+\,{}^4{\rm He} + R'
\label{eq:dc_res}
\end{equation}
described by the rate constant $k_{\rm DC}^g$.  At the MeV ion beam
energies used here, it is expected that $k_{\rm DC}^g\ll k_{\rm
  DE}^{g}$; also, we assume for simplicity that at these collision
energies both $k_{\rm DE}^{g}$ and $k_{\rm DC}^g$ are independent of
the initial rovibrational state of the molecular ion.  The rate
constant $k^g_X$ for any rest-gas induced process $X$ is related to
the collision cross section $\sigma^g_X$ by 
\begin{equation}
k^g_X = n_gv_i\sigma^g_X,
\label{eq:kg_general}
\end{equation}
where $n_g$ ($\sim$10$^6$ cm$^{-3}$ in the present experiments) is the
residual gas density.

The ro-vibrationally averaged rate constants of electron-induced
reactions follow from Eq.\ (\ref{eq:rx_toroid}) as
\begin{equation}
\tilde{k}_X(E_d;t)=\eta n_e \left[ \tilde{\alpha}_X(E_d) 
+ \tilde{\alpha}^{\rm tor}_{X}(E_d) \right].
\label{eq:kx_general}
\end{equation}
Rate constants $k^{vJ}_X(E_d)$ for electron-induced reactions on ions
in a specific ro-vibrational level are defined in a similar manner.
The loss of ions from the circulating beam is then described by the
following rate equation for the ion number $N_i$:
\begin{equation}
\dot{N_i} = -\left[
   2k_{\rm DE}^{g} 
 +  k_{\rm DC}^g         
 +   \tilde{k}_{\rm DR}(E_d;t)
 + 2 \tilde{k}_{\rm DE}(E_d;t)                       
 \right] N_i,
\label{Nion}
\label{eq:master1}
\end{equation}
where rate constants for the electron-induced DR and DE reactions were
introduced.  

Similarly, the count rates representing the arrival of one or the
correlated arrival of two neutral particles at the surface barrier
detector (see Sec.\ \ref{sec:exp_overview}) can be described; the
properties of this detector allow unit counting efficiency to be
assumed.  For events yielding two neutrals ($^3$He + $^4$He) the rate
is given by
\begin{equation}
\rdr = \left[ f_gk_{\rm DC}^g + \tilde{k}_{\rm DR}(E_d;t)\right] N_i,
\label{eq:master2}
\end{equation}
while the rates for the observation of a single neutral fragment
($^3$He or $^4$He) are given by
\begin{equation}
\rdei = \rdeii = \left[ f_gk^g_{\rm DE} + \tilde{k}_{\rm DE}(E_d;t)\right] N_i.
\label{eq:master3}
\end{equation}
Here $f_g$ is a geometrical factor expressing the fraction of all
residual gas events (which occur over all the ring circumference) that
leads to a neutral particle detection on the surface barrier detector.

\subsubsection{Radiative thermalization}
\label{sec:beam_evolution_rad}

\begin{figure}[htbp]
\includegraphics[width=3.0in]{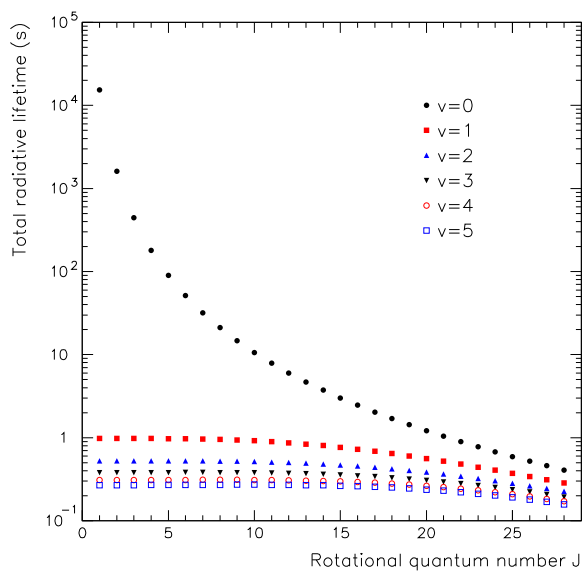}
\centering
\caption{
  Calculated total radiative lifetimes as a function of the rotational
  quantum number $J$ for the first six vibrational levels ($v=0$--5)
  in the $X{^2}\Sigma_u^+$ electronic ground state of
  $^3$He$^4$He$^+$.}
\label{lifetime}
\end{figure}

Since the ${^3}$He${^4}$He$^+$ molecular ions have a permanent
electrical dipole moment, their rovibrational motion is coupled to the
300-K radiation field of the storage ring.  Based on the potential
energy curve \cite{cencek1995} for the $X^{2} \Sigma_u^{+}$ electronic
ground state of He$_2^+$ we have calculated the Einstein coefficients
for spontaneous emission ($A_{f}^{i}$), stimulated emission
($S_{f}^{i}$), and absorption ($B_{f}^{i}$) for rovibrational
transitions connecting initial ($i$) and final ($f$) rovibrational
levels; the same approach and approximations as in Ref.\ 
\cite{amitay1994} were used \cite{amitay1994_note}.  Since the
spin-rotational interaction is negligible, the rotational quantum
numbers are $J=0,1,2,\ldots$ and the linestrengths for
${^1}\Sigma$\,-${^1}\Sigma$ transitions have been used instead of
those for ${^2}\Sigma$\,-${^2}\Sigma$ transitions \cite{herzberg},
with the selection rule $\Delta J = \pm 1$.  Figure \ref{lifetime}
shows the results of these calculations for the total radiative
lifetime $\tau_{vJ} = 1 / \sum_{(v'J')<(vJ)} A_{v'J'}^{vJ}$, as a
function of rotational quantum number $J$ for the first six
vibrational states of the ${^2 \Sigma_u^+}$ ground state of
${^3}$He${^4}$He$^+$. [The notations $(v'J')<(vJ)$ and $(v'J')>(vJ)$
are used to label the levels lying energetically below and above
a given level $vJ$, respectively.]

The master equations for the numbers $N_{vJ}$ of stored ions
in states $vJ$ under the influence of blackbody-induced radiative
transitions are given by
\begin{eqnarray}
\dot{N}_{vJ} & = &\sum_{(v'J')>(vJ)} \left\{\raisebox{0mm}[3mm][3mm]{}
  \left[A_{vJ}^{v'J'}+S_{vJ}^{v'J'}\rho_T(E_{vJ}^{v'J'})\right] N_{v'J'}
  \;-
\right.\nonumber\\
&&\hspace*{1cm}
\left.\raisebox{0mm}[3mm][3mm]{}
  - B_{v'J'}^{vJ}\rho_T(E_{v'J'}^{vJ}) N_{vJ}
\right\} + 
\nonumber\\
&&\sum_{(v'J')<(vJ)} \left\{\raisebox{0mm}[3mm][3mm]{}
    B_{vJ}^{v'J'}\rho_T(E_{vJ}^{v'J'}) N_{v'J'} \; -
\right.\nonumber\\
&&\hspace*{1cm}
\left.\raisebox{0mm}[3mm][3mm]{}
 - \left[A_{v'J'}^{vJ}+S_{v'J'}^{vJ}\rho_T(E_{v'J'}^{vJ})\right] N_{vJ}
\right\},\nonumber\\
\label{modelevolution}
\end{eqnarray}
where $\rho_T(h\nu)$ is the Planck
distribution for radiation at temperature $T$ at photon energy $h\nu=E_{f}^{i}=|E_f-E_i|$.
Master equations of this type, possibly including further, $v$ and $J$
dependent interactions of the stored ions, are used to model the time
evolution of level populations $p_{vJ}(t)$ in the stored ion beam,
defined by $p_{vJ}(t)=N_{vJ}(t)/\sum_{v'J'}N_{v'J'}(t)$.

\begin{figure}[htbp]
\centering
\includegraphics[width=3.0in]{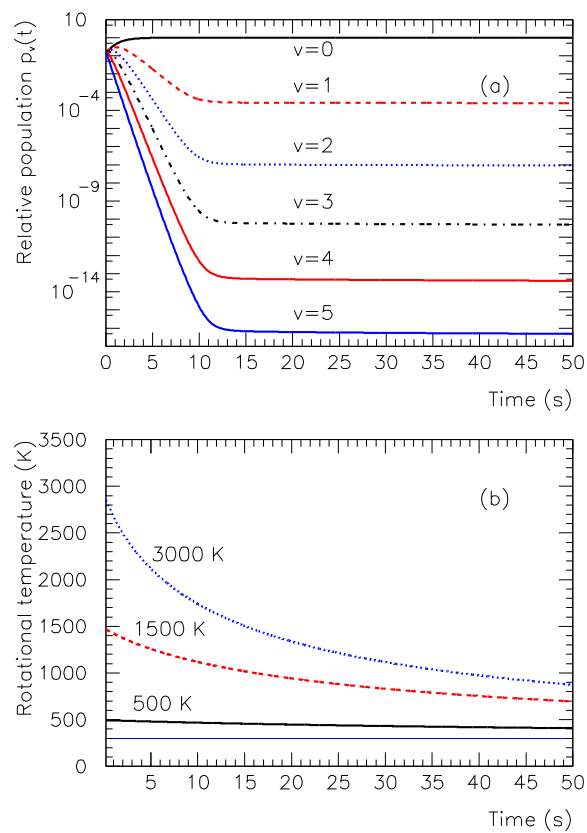}
\caption{
  Radiative thermalization of rovibrational levels in the
  $X{^2}\Sigma_u^+$ electronic ground state of $^3$He$^4$He$^+$.  
  (a) Vibrational thermalization among $v=0$--5.  Initially, the six
  vibrational levels were populated equally, with a rotational
  temperature of 1500 K imposed on each level.  (b) Rotational
  thermalization in the $v=0$ level for initial rotational
  temperatures of 500 K (solid), 1500 K (dashed), and 3000 K
  (dotted). The lowest line marks the 300 K equilibrium 
  temperature.}
\label{radiative1}
\end{figure}

To illustrate the significance of radiative transitions on the
evolution of the rovibrational population of the stored ion beam, and
to predict the associated relaxation times, the coupled set of Eqs.\ 
(\ref{modelevolution}) was solved for a Planck radiation field at 300
K for different initial populations (Fig.\ \ref{radiative1}) using a
fourth order Runge-Kutta method for numerical integration.

Thermalization among the six lowest vibrational levels [Fig.\ 
\ref{radiative1}(a)] occurs after $\sim$12 s; however, already at
$\sim$5 s the populations of higher vibrational levels ($v\ge2$) are
well below $10^{-4}$, and the characteristic time scale for changes in
the vibrational populations amounts to only $\sim$3 s.  The rotational
populations within the $v=0$ level [Fig.\ \ref{radiative1}(b)] evolve
over several tens of seconds, and thermalization is only reached after
hundreds of seconds.  If only radiative transitions were decisive for
the evolution of the rovibrational population in the present
measurements, higher vibrational states ($v\ge3$) would be essentially
unpopulated after $\sim$3 s of storage and no contribution to the DR
signal [Eqs.\ (\ref{eq:alp_mean_general}),(\ref{eq:master2})] should
occur from these states despite a DR cross section four orders of
magnitude larger than for the ground state ($v=0$).

\subsubsection{Collisional effects on rovibrational populations}
\label{sec:beam_evolution_coll}

Collisions with both residual gas molecules and electrons are
known to affect the rovibrational populations in stored beams of
molecular ions \cite{zajfman2003,lammich2003,tanabe1999,krohn2000}.
The relevant elementary reactions between the stored He$_2^+$ ions and
residual gas molecules in the present case would be collisions at MeV
energies leaving the interacting He$_2^+$ intact, but changing its
rovibrational excitation.  In a photodissociation study of the
stronger radiatively active molecular ion CH$^+$
\cite{hechtfischer1998} it was found that the ions do equilibrate with
the 300 K radiation field, and effects of collisional induced
rotational heating (expressed as an increase of the rotational
temperature) were determined to be $\le$10 K/s at similar residual gas
densities as in the present experiment.  In a study of HD$^+$ under
similar conditions, evidence for a weak but measurable effect of
vibrational excitation in the residual gas was reported
\cite{lange_phd,zajfman2003}.

The interaction with the electron beam can affect the rovibrational
populations in two ways.  Firstly, different rate coefficients
$\alpha_X^{vJ}(E_d)$ for electron-induced reactions $X$ (in particular
DR, possibly DE) can lead to different ion loss rate constants
$k_X^{vJ}(E_d)$ [Eq.\ (\ref{eq:kx_general})] for the various
rovibrational levels, leading to {\em preferential depletion} of
levels with higher reaction rates.  Indeed, in a recent study of the
DR of D$_{2}$H$^{+}$ \cite{lammich2003} this was considered as a
mechanism likely to produce rotational populations in the ion beam
with temperatures below the ambient blackbody radiation field.
Secondly, (ro-)vibrational transitions can be directly induced in
electron collisions, as described by Eq.\ 
(\ref{eq:general_ex}).  In particular, vibrational excitation is
likely to occur at collision energies in the eV range, as found for
HD$^+$ \cite{lange_phd,zajfman2003}.  Even if the interaction energy
is tuned to small $E_d$ in the central part of the electron cooler,
higher electron collision energies occur over smaller portions in
the toroid region (see Sec.\ \ref{sec:exp_overview}) and can cause vibrational
excitation.

On the other hand, at vanishing detuning energy $E_d$, where many of
the present studies are performed, collisions in the central region of
the electron cooler are characterized by very small interaction
energies, as given by the electron temperature $kT_\perp = 10$ meV.
Under these conditions, the rate coefficients for superelastic
collisions (SEC) \cite{nakashima1986} with vibrationally excited ions
often exceed the DR rate coefficient, and the related vibrational
de-excitation [cf.\ Eq.\ (\ref{eq:general_ex})] was indeed
experimentally observed for H$_2^+$ \cite{tanabe1999,krohn2000} and
D$_2^+$ \cite{zajfman2003}.

Also in the present experiment on ${^3}$He${^4}$He$^+$ the electron
temperature $kT_\perp$ is well below the vibrational spacings, and of
the order of the rotational spacings for low-lying $J$ levels in the
$v=0$ state.  As purely rotational radiative transitions are
relatively slow [see Figs.\ \ref{lifetime}, \ref{radiative1}(b)], {\em
  rotationally} inelastic electron collisions may compete with them.
With a rotational constant of $B_e=8.4$ cm$^{-1}$ $\approx$ 1 meV
\cite{yu1987} for ${^3}$He${^4}$He$^+$, both excitation and
de-excitation of rotational levels can be caused by electron
collisions.  The rates of rotationally inelastic low-energy
electron-ion collisions have been considered theoretically for several
molecular systems (see for instance Refs.\ 
\cite{rabadan1998a,rabadan1998b,faure2001} and references therein),
however, to the best of our knowledge no calculations exist for
${^3}$He${^4}$He$^+$.  The typical cross sections for these reactions
are in the range of 10$^{-14}$--10$^{-12}$ cm$^2$ near their thresholds, and
their effect on the present measurement can indeed be significant.

Collisional effects on the rovibrational populations in the stored ion
beam will be further discussed below in Sec.\ \ref{sec:modelcalc_rot},
considering both residual gas and electron collisions.  The radiative rates of
Eq.\ (\ref{modelevolution}) will be complemented by approximate rate
constants for inelastic collisions in order to obtain a rough model
description of the rovibrational evolution under such effects.  It
should be emphasized that the evolution of the populations cannot generally
be slowed down, but only accelerated by the occurrence of additional
processes.  Hence the {\em maximum} timescales over which the
vibrational and rotational level populations evolve will still be
given by the radiative timescales determined above.

\section{Coulomb explosion diagnostics}
\label{sec:exp_CEI}
\label{sec:vibrationalpopulation}

The relative vibrational level populations of the stored
${^3}$He${^4}$He$^{+}$ ions were monitored with Coulomb explosion
imaging (CEI) of ions slowly extracted from the beam during storage.
Details about the CEI method and the setup at the TSR with the slow
ion extraction can be found in Refs.\ \cite{wester1998,vager1989,%
  zajfman1994}.  The extracted ions with kinetic energy $E_i=
(M/2)v_i^2 = 7.28$ MeV were collimated by two 1-mm apertures 3 m
apart, and then sent through a 80 \AA\ thick diamond like carbon foil;
the thickness corresponds to a dwell time of the ions in the foil of
only $5.6 \times 10^{-16}$ s.  In the foil the binding electrons are
rapidly (within $\sim$10$^{-16}$ s) stripped from the molecule,
leaving the bare nuclei to separate (explode) due to their mutual
Coulomb repulsion and thus to convert their Coulomb energy into
kinetic energy.  The stripping time is much faster than both the
vibrational ($\tau_{\rm vib} = 2/c \omega_e = 4 \times 10^{-14}$ s
\cite{yu1987}) and rotational periods ($\tau_{\rm rot} = 1/2cB_e = 2
\times 10^{-12}$ s \cite{yu1987}) of ${^3}$He${^4}$He$^+$; so the
nuclear motion is essentially unchanged prior to the explosion, and to
a first approximation the kinetic energy release therefore reflects
directly the incident molecule's nuclear coordinates.  However,
despite the short dwell time in the foil, the recoiling nuclei also
interact with the target atoms \cite{zajfman1992,garcia-molina2000}
which causes an alteration of the kinetic energy release from the
value expected for a pure Coulomb explosion.

The actual kinetic energy release $\tilde{E_k}$ was determined at a
distance $\tilde{s}=2.965$ m from the foil with a three dimensional
imaging technique \cite{wester1998}, where the relative distance
$\tilde{D}$ on the plane of the detector and the relative time of
arrival $\Delta t$ of the two fragments from an exploding molecule
were recorded.  The position resolution was $\sigma_{\tilde{D}}= 0.1$
mm and the time resolution was measured to be $\sigma_{\Delta t}=100$
ps (1$\sigma$ Gaussian widths).  From the time and position data, the
kinetic energy release is determined as
\begin{equation}
  \tilde{E_k}=\frac{E_i}{\tilde{s}^2} 
    \, \frac{ M(^3{\rm He})M(^4{\rm He}) } {[M(^3{\rm He})+M(^4{\rm He})]^2}
    \, [(v_i \Delta t)^2 + \tilde{D}^2].
\label{Ecoulomb}
\end{equation}
The kinetic energy distribution after Coulomb explosion for an
ensemble of ions in a given vibrational state $v$ has a characteristic
form $P_v(\tilde{E_k})$ \cite{amitay1999} that reflects the
distribution of nuclear positions and momenta in this state as well as
the fragment scattering in the foil, broadening the distribution by
$\sim$20\%.  For an ensemble of ions in several vibrational states the
normalized kinetic energy distribution after Coulomb explosion 
\begin{equation}                   
      P(\tilde{E_k},t)=\sum_v p_v(t) P_v(\tilde{E_k}) 
\label{Pcoulomb}
\end{equation}
is a superposition of the normalized distributions for the individual
vibrational states, where the coefficients $p_v(t)$ represent the
relative populations of the vibrational states $v$ in the beam, and
$\sum_v p_v(t) = 1$.

The resolution of the CEI technique does not allow us to distinguish
small contributions to the kinetic energy from rotational excitation,
and the coefficients are to be considered summed over the rotational
degrees of freedom
\begin{equation}  
        p_v(t)=\sum_J p_{vJ}(t).
\label{population}
\end{equation}
With an accurate modeling of the distributions $P_v(\tilde{E_k})$ the
relative distribution of vibrational states in the beam can be
determined by fitting Eq.\ (\ref{Pcoulomb}) to the experimental
distribution with the relative populations $p_v(t)$ as free
parameters.

To model the distributions $P_v(\tilde{E_k})$ we followed closely the
procedure described in Ref.\ \cite{amitay1999}, where a quantum
mechanical description of the Coulomb explosion process was combined
with semi-classical ion propagation through the foil.  Briefly, the
nuclear wave functions were first calculated by numerically solving
the radial nuclear Schr\"{o}dinger equation for $^3$He$^4$He$^+$ in
the electronic potential given in Ref.\ \cite{cencek1995}.  Each
nuclear wavefunction was projected on the Coulombic continuum of
$^3$He$^{2+}$ and $^4$He$^{2+}$ using Coulomb wave functions
calculated with the WCLBES subroutine \cite{thompson1985,thompson1986}
available from the CERN Fortran Library, to obtain the kinetic energy
distribution corresponding to Coulomb explosion without scattering
effects in the foil.  This kinetic energy distribution was inverted to
an artificial distribution of internuclear distances ($R$) by the
transformation $\tilde{E}_k = 4e^2/R$.  

For classical trajectories in the Coulomb explosion, this artificial
radial distribution yields the distribution of kinetic energy releases
that corresponds to the quantum mechanical Coulomb explosion.  The
artificial distribution of initial radial distances $R$ was then used
as input to a semi-classical propagation of ions through the carbon
foil, which modeled multiple scattering and charge exchange of the
molecular fragments in the foil \cite{zajfman1992} as well as the
effects of wake fields and electronic stopping
\cite{garcia-molina2000,lammich2004}.  After leaving the foil the ions
were propagated on classical trajectories to the detector, where also
the calculated position and times were folded with the finite
resolutions ($\sigma_{\tilde{D}}$ and $\sigma_{\Delta t}$) of the
detector.  The semi-classical propagation as well as the folding of
the simulations with the detector characteristics were done using
existing codes \cite{zajfman1992,lammich2004}.

\begin{figure}[htbp]
\centering
\includegraphics[width=3.0in]{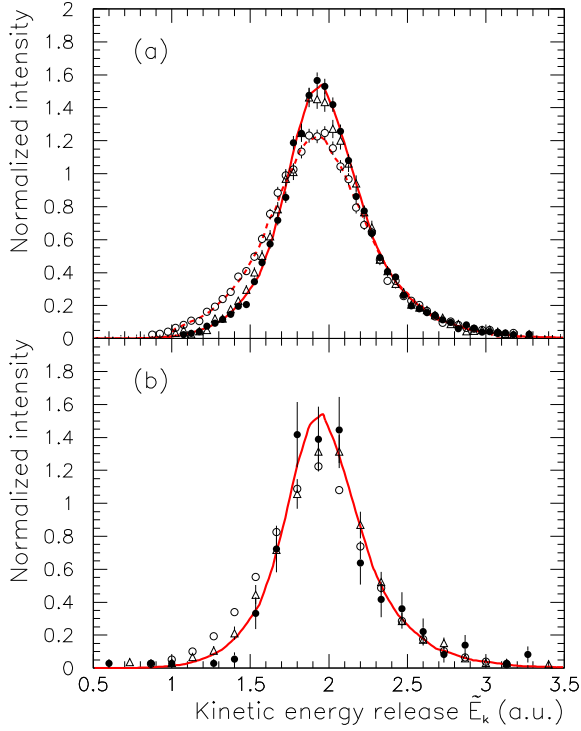}
\caption{
  Normalized distributions of the kinetic energy releases
  ($\tilde{E}_k$) after foil induced Coulomb explosion of
  $^3$He$^4$He$^+$ ions extracted from the stored beam. (a) Electron
  beam off; (b) electron beam on ($E_d=0$).  Distributions are shown
  for times after injection of 0--1 s (open circles), 2--3 s
  (triangles), and $\ge$3 s (filled circles).  The solid lines show
  the simulated distribution for the ground state, and the dashed line
  shows the result of a least-squares fit with Eq.\ (\ref{Pcoulomb})
  to the distribution without the electron beam obtained at 0--1 s.
  Ion energy $E_i=7.28$ MeV; electron density $n_e=5.5\times10^6$
  cm$^{-3}$.}
\label{cei_data}
\end{figure}

Figure \ref{cei_data}(a) displays normalized kinetic energy
distributions $P(\tilde{E}_k,t)$ as measured with the CEI technique
for three time intervals after ion injection into the TSR when no
electron cooling was applied.  The kinetic energy distribution is seen
to become narrower with time, and the distribution seems to
converge after 3 s, i.e., no change of the distribution could be
observed when comparing distributions at different time intervals for
$\ge$3 s after injection.  When no electron cooling is applied the
rovibrational population of the ion beam is determined by radiative
thermalization in competition with excitation through ion collisions
with the residual gas (see Sec.\ \ref{sec:beam_evolution_coll}).  The
relaxed kinetic energy distribution ($t\ge 3$ s) without electron
cooling compares very well with the Monte Carlo simulation for
$P_{v=0}(\tilde{E}_k)$, obtained from the procedure described above.
Performing a least-squares fit to the same distribution, using the
superposition of simulated functions $P_v(\tilde{E}_k)$ according to
Eq.\ (\ref{Pcoulomb}), yields $p_{v=0}(t\ge 3\,{\rm s}) = (98.9 \pm
1.3)$\%.  For the kinetic energy distribution in an early phase of the
vibrational relaxation ($t=0$--1 s) the level populations are found
from a similar fit to be $(56 \pm 3)$\% in $v=0$, $(13 \pm 4)$\% in
$v=1$, $(16\pm 5)$\% in $v=2$, and 15\% distributed over the levels
with $v \ge 3$.

Figure \ref{cei_data}(b) displays the normalized kinetic energy
distributions $P(\tilde{E}_k,t)$ for the same time intervals as in
Fig.\ \ref{cei_data}(a) with electron cooling applied continuously at
all storage times, in parallel with the slow extraction.  
The combined action of the electron beam and the slow extraction 
strongly reduces the beam lifetime and the amount of
data is rather low.  However, also in this case the kinetic energy
distribution is seen to narrow with time, and after 3 s the
experimental distribution compares well with the simulated one for the
ground state.  A least-squares fit to the converged distribution,
using the superposition of simulated functions $P_v(\tilde{E}_k)$ as
given in Eq.\ (\ref{Pcoulomb}), yields $p_{v=0}=$ 91 $\pm$ 14 $\%$.

Summarizing, from the CEI measurements it is evident that the ion beam
is dominated by the vibrational ground state $v=0$ after 3 s both with
and without the presence of the electron beam during storage.  This is
consistent with the time scale of radiative thermalization, but both 
thermal or non-thermal vibrational distributions with a few percent of 
the ions in higher vibrational states are consistent with the data.

\section{Electron-induced reactions}
\label{sec:drbasic}

\subsection{DR and DE rate measurements}
\label{sec:dr-rate}

Details of the experimental procedure for measuring rates of fragments
from DE and DR reactions at the TSR have been described previously
\cite{kilgus1992,amitay1996}.  In the present measurement, the rates
of neutral fragment events with total masses of 3, 4 and 7 atomic mass
units were observed with the surface barrier detector (Sec.\ 
\ref{sec:exp_overview}) as a function of the electron detuning energy
$E_d$ and the time $t$ after the injection, yielding the quantities
$\rdr(E_d;t)$, $\rdei(E_d;t)$, and $\rdeii(E_d;t)$, respectively.  The
electron detuning energy could be cycled between the cooling energy
($E_d=E_d^{c}=0$), a variable ``measurement'' energy $E_d^m$, and a
``reference'' energy $E_d^r$.  After choosing $E_d^r$ appropriately as
described below, the full-mass count rate at the reference energy,
$\rdr(E_d^r;t)$, served for normalization with respect to the stored
$^3$He$^4$He$^+$ ion current.

The energy dependences of the count rates $R$ were
usually obtained by cycling (``wobbling'') between the three
given levels of $E_d$ at a dwell time of 50 ms on each of the levels.
Following an injection of ions into the ring, $E_d$ was first kept
constant at $E_d^{c}=0$ for 5 s to allow for phase space cooling and
vibrational relaxation of the stored ion beam whereafter the electron
energies $E_d$ were wobbled as described.  The value $E_d^m$ was
varied from one injection to the next.  Energy and time dependent
rates $\rdr(E_d^m;t)$, $\rdei(E_d^m;t)$, and $\rdeii(E_d^m;t)$ were
obtained from the detector counts accumulated over numerous
injections, performing several scans of $E_d^m$ over the desired
range, where the counts for individual values of $E_d^m$ and for
individual bins of the ion storage time (i.e., the time after the
injection into the storage ring) were added.  
With normalization and background subtraction as described below, rate
coefficients $\tilde{\alpha}_{\rm DR}(E_d;t)$ and $\tilde{\alpha}_{\rm
  DE}(E_d;t)$ for the electron induced reactions were obtained.  Time
variations in these rate coefficients can arise from variations of the
$^3$He$^4$He$^+$ level populations $p_{vJ}(t)$ in the stored beam [see
Eq.\ (\ref{eq:alp_mean_general})].

Apart from the wobble scheme, also much longer dwell times for $E_d$,
ranging up to many seconds, were used in order to study the temporal
behavior of the absolute and the normalized count rates, aiming at
the determination of absolute rate coefficients and at the study of
the level populations $p_{vJ}(t)$ of the stored ions.  Measurements
with both types of timing schemes will be discussed in the following.

\subsubsection{Relative, energy dependent count rates}

\begin{figure}[htbp]
\centering
\includegraphics[width=3.0in]{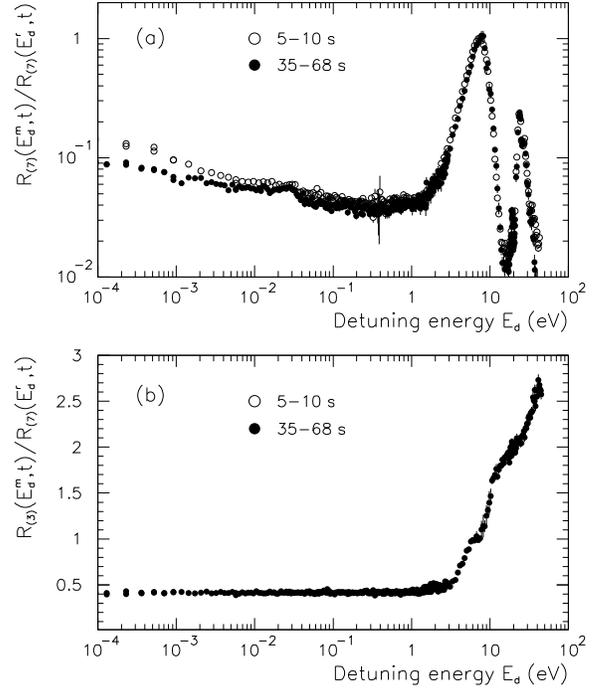}
\caption{Measured rates (raw data) of (a) DR and (b) DE events 
with normalization to the rate of DR events at the reference 
detuning energy ($E_d^r$=7.3 eV).
Ion energy $E_i=7.28$ MeV; electron density $n_e=5.5\times10^6$ cm$^{-3}$.}
\label{fig:raw}
\end{figure}

Typical raw data for the DR and the DE channel, 
averaged over time intervals of 5-10 s and 35-68 s and
representing the relative rates 
$\rdr(E_d^m;t)/\rdr(E_d^r;t)$ and $\rdei(E_d^m;t)/\rdr(E_d^r;t)$, 
respectively, are shown in Fig.\ \ref{fig:raw}.  
In these data, as in the remainder of this paper, the
reference energy was chosen as that of the prominent peak in the DR
signal $\rdr$, setting $E_d^r=7.3$ eV.  The relative DR rate
coefficient shown in Fig.\ \ref{fig:raw}(a) shows a considerable
storage time dependence at $E_d\lesssim 0.1$ eV, while the high-energy
structure does not vary significantly with the storage time.  The
absolute DR rate coefficient at $E_d^r$ and the rate variations with
the storage time will be considered further in the following.  The DE
rate in Fig.\ \ref{fig:raw}(b) shows a threshold at $\sim$2.5 eV,
corresponding to the dissociation energy of He$_2^+$.  The non-zero rate
$\rdei$ below the threshold arises partly from DE events in the
residual gas and partly from the toroid contribution discussed in Sec.\ 
\ref{sec:rate_coeff}.

\subsubsection{Beam lifetime, reaction rate constants, and absolute
  rate coefficients}
\label{sec:drtime}
                
A series of measurements with long dwell times on different levels of
$E_d$ was performed in order to study the influence of electron and
rest-gas induced reactions on the ion beam lifetime, to clarify the
relative significance of the various reaction rates considered in
Sec.\ \ref{sec:beam_loss}, and to obtain first information about the
effects of time-varying internal excitations, as represented by
$p_{vJ}(t)$.  Moreover, these measurements yield an absolute
calibration of our rate coefficients.

\begin{figure}[htbp]
\centering
\includegraphics[width=3.0in]{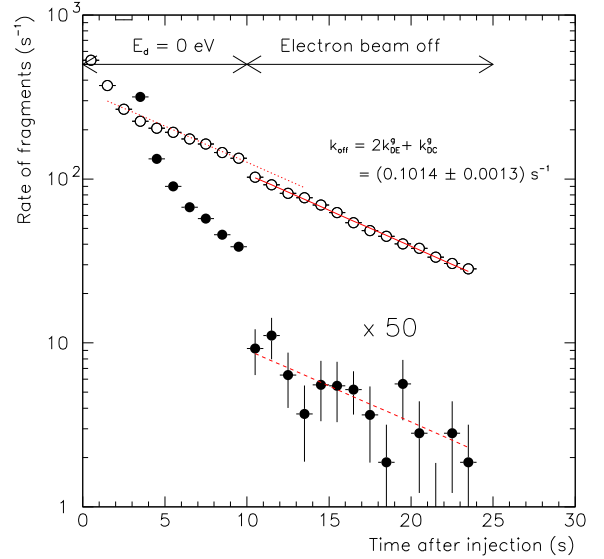}
\caption{
  Measured fragment rates $\rdr$ (filled circles) and $\rdei$ (open
  circles) averaged over 76 injections when the electron beam is
  switched off after 10 s of storage.  For times $>$10 s the values
  $\rdr$ (solid circles) have been multiplied by a factor of 50. 
  The solid line shows a single exponential fit to the rate $\rdei$ at 
  10--23 s with the electron beam off, while the dashed lines compares 
  the slope of the fitted curve to the rate $\rdei$ at ealier times 
  and to the  rate $\rdr$. Ion energy $E_i=7.28$ MeV; electron density 
  $n_e=5.5\times10^6$ cm$^{-3}$.}
\label{electron_off}
\end{figure}

Figure \ref{electron_off} shows the time evolution of the absolute
rates $\rdr$ and $\rdei$ for a situation where the electron beam was
continuously on at $E_d=0$ for 10 s, whereafter it was switched off.
The transient behavior observed in $\rdei$ at $<5$ s can be attributed to 
changes in the ion beam orbit and hence its pointing towards 
the detector during the initial electron cooling of the ion beam.

The full-mass rate $\rdr$ shows a clearly non-exponential decrease up
to the time when the electron beam is turned off, when it drops
sharply, by more than two orders of magnitude.  We conclude that this rate is
dominated by DR and that the related rate constant $\tilde{k}_{\rm
  DR}(0;t)$ shows a significant time dependence, representing that of
$p_{vJ}(t)$, which will be further investigated below.  
On the other hand, at times later than 5 s the decrease of $\rdei$ 
is well described by a single exponential, both with and without 
the electrons. 
The jump in $\rdei$
at the time when the electron beam is switched off indicates the
presence of an electron-induced DE rate even at $E_d=0$, representing
the toroid contribution discussed above; this rate is represented by
the term $\tilde{k}_{\rm DE}(0;t)$ in Eq.\ (\ref{eq:master3}).  From
the observed step in $\rdei$ at $t=10$ s, we find the size of this
contribution relative to the rest-gas rate constant as
\begin{equation}
c_1=\tilde{k}_{\rm DE}(0) / f_g k_{\rm DE}^g
= 0.18(2).
\label{eq:c1}
\end{equation}
No significant time dependence of $\tilde{k}_{\rm DE}(0)$ is observed,
as $\rdei$ decreases as a single exponential even with the electrons
present.

The time dependence of $\rdei$ at $t>10$ s reveals the beam loss due
to rest-gas induced processes, described by $\dot{N}_i(t)=-k_{\rm off}
N_i(t)$ with $k_{\rm off} = 2k_{\rm DE}^g + k_{\rm DC}^g$ [cf.\ Eq.\ 
(\ref{eq:master1})].  Dissociative charge exchange (DC) in the rest gas is
much less likely than DE, as revealed by the ratio $\rdr/\rdei$
without electron beam, which yields with the help of Eqs.\ 
(\ref{eq:master2}) and (\ref{eq:master3})
\begin{equation}
c_2 = k_{\rm DC}^g / k_{\rm DE}^g = 1.6(2) \times 10^{-3}.
\label{eq:c2}
\end{equation}
From the experimental result $k_{\rm off} = 0.1014(13)$ s$^{-1}$ one
can then deduce
\begin{equation}
k_{\rm DE}^g = k_{\rm off} /(2+c_2)= 0.0506(7)~{\rm s}^{-1}.
\label{eq:kgde}
\end{equation}
together with $k_{\rm DC}^g=8(1)\times10^{-5}$ s$^{-1}$.

\begin{figure}[htbp]
\centering
\includegraphics[width=3.0in]{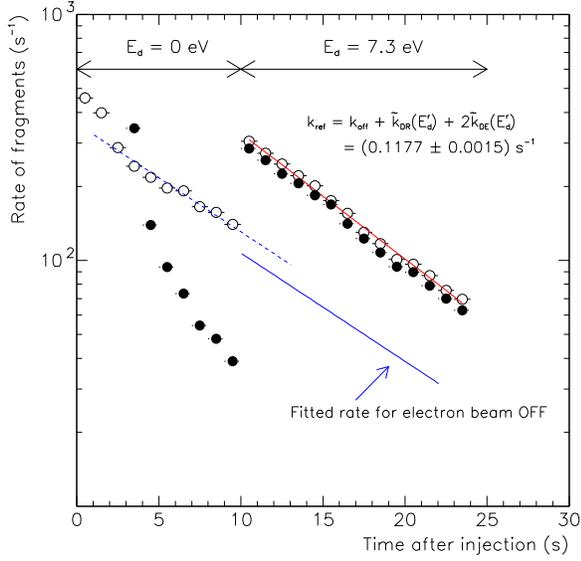}
\caption{
  Measured rates $\rdr$ (filled circles) and $\rdei$ (open circles)
  average over 33 injections as a function of time $t$ after injection
  where the detuning energy $E_d$ was changed from 0 to 7.3 eV at
  $t=10$ s. The upper solid line shows a single exponential fit to 
  the rate $\rdei$ at 10--23 s, while the lower dashed and soild 
  lines are repeated from Fig. \ref{electron_off}.	}
\label{electron_73ev}
\end{figure}

The measurement of Fig.\ \ref{electron_off} was repeated immediately,
under the same vacuum conditions, now changing the electron energy
$E_d$ from 0 to $E_d^r=7.3$ eV at $t=10$ s (see Fig.\ 
\ref{electron_73ev}).  The full-mass rate $\rdr$, showing the same
non-exponential behavior as before for $E_d=0$, increases strongly as
$E_d$ is stepped up to $E_d^r$ and turns over to a single-exponential
decay.  The decay rate of $\rdr$ for $t>10$ s is the same as that of
$\rdei$ within the experimental accuracy and amounts to $k_{\rm ref} =
0.1177(15)$ s$^{-1}$.

The single-exponential decays of $\rdr$ and $\rdei$ at $E_d^r$
indicate that $\tilde{k}_{\rm DR}(E_d^r)$ and $\tilde{k}_{\rm DE}(E_d^r)$
are time independent, i.e., they reveal no influence due to possible
time variations of the level populations $p_{vJ}(t)$.  Using Eqs.\ 
(\ref{eq:master2}), (\ref{eq:master3}), (\ref{eq:c1}), and
(\ref{eq:c2}) to express the measured ratio
$c_3=\rdr(E_d^r)/\rdei(0)=2.56(7)$ from Fig.\ \ref{electron_73ev}, one
finds
\begin{equation}
\frac{\tilde{k}_{\rm DR}(E_d^r)}{f_g k_{\rm DE}^g} =
c_3(1+c_1)-c_2
\label{eq:frac1}
\end{equation}
Using this result and expressing in a similar way the measured ratio
$c_4=\rdei(E_d^r)/\rdr(E_d^r)=1.02(1)$ from Fig.\ \ref{electron_73ev}, 
one finds that the ratio of the DE and DR rate constants at the reference
energy can be expressed as
\begin{equation}
\rho_{\rm DE}=\frac{\tilde{k}_{\rm DE}(E_d^r)}{\tilde{k}_{\rm DR}(E_d^r)} = 
\frac{c_3c_4(1+c_1)-1}{c_3(1+c_1)-c_2}
\label{eq:frac2}
\end{equation}
Evaluation of this expression yields $\rho_{\rm DE}=0.690(14)$.
The decays of both $\rdr$ and $\rdei$ finally
reflect the decrease of the ion number according to
$\dot{N}_i(t)=-k_{\rm ref} N_i(t)$ with 
$
k_{\rm ref} = k_{\rm off} +
\tilde{k}_{\rm DR}(E_d^r)+2\tilde{k}_{\rm DE}(E_d^r)
$ [cf.\ Eq.\ (\ref{eq:master1})].  With the help of Eq.\ 
(\ref{eq:frac2}) it is then possible to obtain the absolute DR rate
coefficient at $E_d^r$ as
\begin{equation}
\tilde{\alpha}^{{\rm c}+{\rm tor}}_{\rm DR}(E_d^r) = \frac{\tilde{k}_{\rm DR}(E_d^r)}{\eta n_e} = 
\frac{k_{\rm ref}- k_{\rm off}}
{\eta n_e[1+2\rho_{\rm DE}]}.
\label{eq:kdr-absolute}
\end{equation}
Here, $\tilde{\alpha}^{{\rm c}+{\rm tor}}_{\rm DR}(E_d^r)$ is used to
denote the sum of the contributions from the central and the toroid
sections, occuring in Eq.\ (\ref{eq:rx_toroid}); a toroid correction
(which is small at $E_d=E_d^r$) will be applied in Sec.\ 
\ref{sec:drenergy}.  With the known values of $n_e(E_d^r)=7.9\times10^6$
cm$^{-3}$ and $\eta=0.027$ we obtain $\tilde{\alpha}^{{\rm c}+{\rm
    tor}}_{\rm DR}(E_d^r)=3.2(4)\times10^{-8}$ cm$^{3}$\,s$^{-1}$, in the 
particular example described above.

One also obtains, with the known DR rate constant $\tilde{k}_{\rm
  DR}(E_d^r)$ and the absolute rate constant $k_{\rm DE}^g$ for DE on
rest-gas from Eq.\ (\ref{eq:kgde}), the efficiency $f_g$ for detecting
rest-gas events, which results in 0.044(7) and hence in an
``observed'' effective ion beam length of $\sim$2.4 m, in reasonable
agreement with the geometrical conditions of the experiment.

The above described procedure to obtain an absolute DR rate
coefficient relies on the assumption that the presence of the electron
beam does not affect the vaccuum significantly in the region of the
electron cooler, i.e., that the values of $k_{\rm DC}^g$ and $k_{\rm
  DE}^g$ are unaffected by the presence of the electron beam.  To
investigate the validity of this assumption we made measurements
similar to the ones in Figs.\ \ref{electron_off} and
\ref{electron_73ev}, but now changing the detuning energy from zero to
0.5 eV at 10 s.  As can be derived from Fig.\ \ref{fig:raw} the
electron induced reactions at 0.5 eV are relatively small, and the
decay of the stored ion beam is determined mainly by DE in the
residual gas ($k_{\rm DE}^g$).  With $E_d=0.5$ eV, an ion beam decay
rate of $k_{\rm 0.5\,eV} = 0.0994(15)$ s$^{-1}$ was obtained, in
consistency with the value $k_{\rm off}$ obtained without the 
electron beam.  Hence,
the electron beam does not affect the beam decay significantly through
an increase of vacuum pressure, and the assumptions underlying the
above derivation are indeed justified.

Repeated measurements of the rate coefficient at $E_d^r$
as described above yielded values that scattered around 
those obtained in the example discussed. 
As the final normalization of the measured rate coefficient 
we use the weigthed average of the different measurements and 
give their standard deviation as the uncertainty, yielding
$\tilde{\alpha}^{{\rm c}+{\rm tor}}_{\rm DR}(E_d^r)
=2.8(4)\times10^{-8}$ cm$^{3}$\,s$^{-1}$.

Altogether, the time dependences of the absolute count rates show a
markedly non-exponential decay for the full-mass events near zero
energy, indicating an influence of internal relaxation of the stored
ions on their DR rate.  For high-energy DR, no such influence is
observed, and a comparison of the decay rates with and without
electrons can be used to obtain the DR rate coefficient on the
peak at 7.3 eV, providing a useful absolute normalization.

\subsubsection{Energy dependent rate coefficients} 
\label{sec:drenergy}    

\begin{figure}[htbp]
\centering
\includegraphics[width=3.0in]{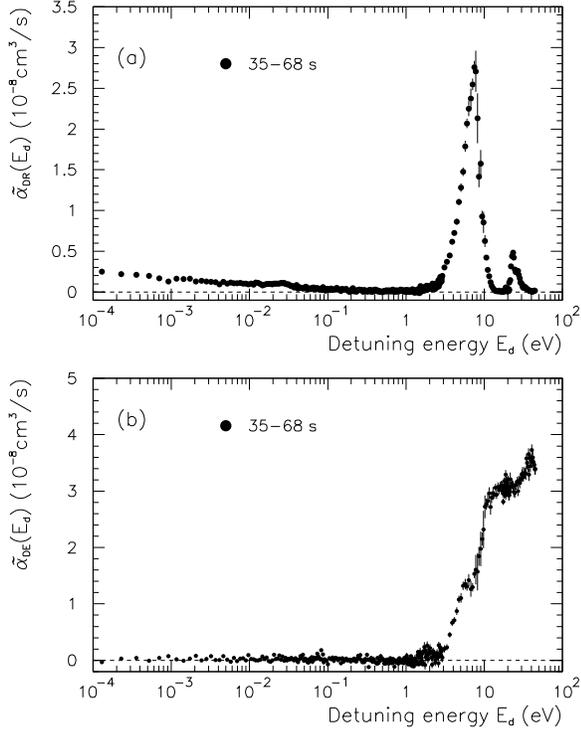}
\caption{
  Rate coefficients as a function of the detuning energy $E_d$
  measured in a time interval of 35--68 s after ion injection.  (a)
  Toroid corrected DR rate coefficient over the full energy
  range studied.  The dashed line marks the zero on the vertical scale.  (b)
  Toroid corrected DE rate coefficient.  Ion energy $E_i=7.28$ MeV;
  electron density $n_e=5.5\times10^6$ cm$^{-3}$.}
\label{rate_dr_de}
\end{figure}

\begin{figure}[htbp]
\centering
\includegraphics[width=3.0in]{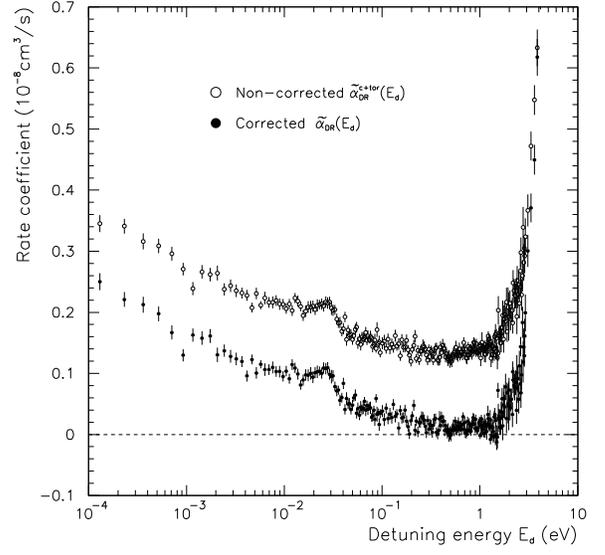}
\caption{
  Effect of the toroid correction on the DR rate coefficient displayed
  in Fig.\ \ref{rate_dr_de}(a) in the low energy region.  The upper
  curve (open circles) shows the directly obtained rate coefficient
  $\tilde{\alpha}^{{\rm c}+{\rm tor}}_{\rm DR}(E_d)$ while the lower
  curve (filled circles) shows the rate coefficient
  $\tilde{\alpha}_{\rm DR}(E_d)$ after toroid correction.}
\label{toroid}
\end{figure}

From the relative rate shown in Fig.\ 
\ref{fig:raw}, the absolute DR rate coefficients can now be
determined using the relation
\begin{equation}
\tilde{\alpha}^{{\rm c}+{\rm tor}}_{\rm DR}(E_d;t)
= \tilde{\alpha}^{{\rm c}+{\rm tor}}_{\rm DR}(E_d^r) \;
\frac{\rdr(E_d;t)}{\rdr(E_d^r;t)}\,
\frac{n_e(E_d^r)}{n_e(E_d)}.
\label{eq:dr-cal}
\end{equation}
Here, the very small background from DC in the residual gas contained
in the rate $\rdr$ can be safely neglected.  Since the DR rate
coefficient for energies outside the scanned energy region can be
assumed to vanish, the standard iterative procedure
\cite{lampert1996,alkhalili2003} can be used to obtain the
toroid-corrected DR rate $\tilde{\alpha}_{\rm DR}(E_d;t)$, shown for
the relaxed beam ($t=35$--68 s) in Fig.\ \ref{rate_dr_de}(a).  The
toroid contribution to the low energy part of the DR rate coefficient
measurement is illustrated by Fig.\ \ref{toroid}.  Its size relative
to the corrected rate coefficient at $E_d=0$ is found to be
$\tilde{\alpha}^{{\rm tor}}_{\rm DR}(0)/ \tilde{\alpha}_{\rm
  DR}(0) = 0.34(6)$.

For DE, the rate coefficient is obtained as 
\begin{eqnarray}
\label{eq:de-cal}
\tilde{\alpha}^{{\rm c}+{\rm tor}}_{\rm DE}(E_d;t)
&=&  \tilde{\alpha}^{{\rm c}+{\rm tor}}_{\rm DR}(E_d^r)
\frac{n_e(E_d^r)}{n_e(E_d)} \; \times \\
&&\times
\frac{\rdei(E_d;t) - \rdei(0;t)(1-\varepsilon_{\rm tor})}
     {\rdr(E_d^r;t)},    \nonumber
\end{eqnarray}
where the background contribution to  DE  from the residual gas only
is subtracted, i.e.\ 
$\varepsilon_{\rm tor}= \tilde{k}_{\rm DE}(0) / 
[f_g k_{\rm DE}^g+\tilde{k}_{\rm DE}(0)]$.
During the experimental runs, small temporal variations of the 
vacuum conditions produce changes in the residual gas rate constant
$k_{\rm DE}^g$.
For the particular residual gas pressure present when the
data used to derive $c_1$ in Eq.\ (\ref{eq:c1}) was taken, we obtain 
$\varepsilon'_{\rm tor}= c_1/(1+c_1)=0.15(2)$.
Since $\tilde{k}_{\rm DE}(0)$ is vacuum independent, the 
value of $\varepsilon_{\rm tor}$ can directly be obtained as
$\varepsilon_{\rm tor}=\varepsilon'_{\rm tor}\,
[\rdei'(0;t)/\rdr'(E_d^r;t)]/[\rdei(0;t)/\rdr(E_d^r;t)]$,
where  $\rdei'(0;t)$ and $\rdr'(E_d^r;t)$
are the observed rates of Figs.\ \ref{electron_off} and 
\ref{electron_73ev}.

The toroid correction of the DE rate is expected to 
be reliable up to $E_d\sim 25$ eV. Above this value an assumption
about the further energy dependence of the DE rate coefficient
from 40 eV up to $\sim 55$ eV is required. We assume as a resonable
approximation that the rate coefficient stays constant in this 
range at the value measured at 40 eV.
The DE rate coefficient corrected under this assumption,
$\tilde{\alpha}_{\rm DE}(E_d;t)$, is shown in 
Fig.\ \ref{rate_dr_de}(b) for $t=35$--68\,s. 
It should be noted that the toroid correction brings the DE rate 
at low energy to zero. Indeed, this verifies in a sensitive manner 
that the identification of the step in the rate $\rdei$ of 
Fig. \ref{electron_off} is correct.

\subsection{DR fragment imaging}
\label{sec:dr_imaging}

The event-triggered fragment imaging detection system mentioned in
Sec.\ \ref{sec:exp_overview} yields the transverse positions of
correlated pairs of neutral $^3$He and $^4$He fragments released in
single DR reactions.  Distributions of the transverse distance between
such correlated product atoms reveal the kinetic energy release (KER)
of the DR events.  The KER in turn can provide a signature of the
internal excitation present in the reacting $^3$He$^4$He$^+$ ions.

Fragment imaging measurements were performed only at $E_d=0$ and
served particularly to clarify the effect of internal excitations on
the low-energy DR rate coefficient $\tilde{\alpha}_{\rm DR}(0;t)$,
found to display a significant time depencence in the measurements of
Sec.\ \ref{sec:dr-rate}.  Here, we will first consider the
experimental framework, analysis, and basic observation related to
fragment imaging of $^3$He$^4$He$^+$ DR reactions.
Similar to the rate coefficient $\tilde{\alpha}_{\rm DR}(0;t)$, also
the fragment imaging distributions show a time dependence, whose basic
phenomenology will be presented in this section.  In a next step
(Sec.\ \ref{sec:rovib_clarify}) we will describe series of
dedicated experiments where both rate coefficient and fragment imaging
measurements are used as tools for understanding the nature and 
the origin of the rovibrational excitation in the stored 
$^3$He$^4$He$^+$ ion beam.

\subsubsection{Fragment imaging experimental conditions}
\label{sec:imaging-conditions}

Following a reaction of $^3$He$^4$He$^+$ with the residual gas or with
electrons, either one or two neutral fragments arrive at the detector,
which for unit detection efficiency and a perfect single-event trigger
scheme would yield corresponding one-body (single) or two-body
(double) events.  The {\em observed} numbers $\tilde{N}_s$ and
$\tilde{N}_d$ are reduced with respect to the true event numbers $N_s$
and $N_d$ due to the finite detection efficiency of the MCP detector.

As the fragments are identified by the light spots they generate on the
detector screen, a two-body event with a small transverse distance can in
addition be falsely identified as a one-body event, in case the light spots
overlap. This results in a limited detection for DR events with small
KER.  

The measured fragment rates were kept below 10$^3$ s$^{-1}$ in all runs.
The ratio of detected double to single events was found to vary
between 0.15 and 0.03.  With an effective coincidence time window of
the event trigger scheme of $\sim$10 $\mu$s, the false coincidence
contribution is expected to stay below 1\%.  Accordingly, the ratio of
(false) triple events to the doubles was found to be well below 1\%.
We hence consider that false coincidences can be neglected.

\begin{figure}[htbp]
\centering
\includegraphics[width=3.0in]{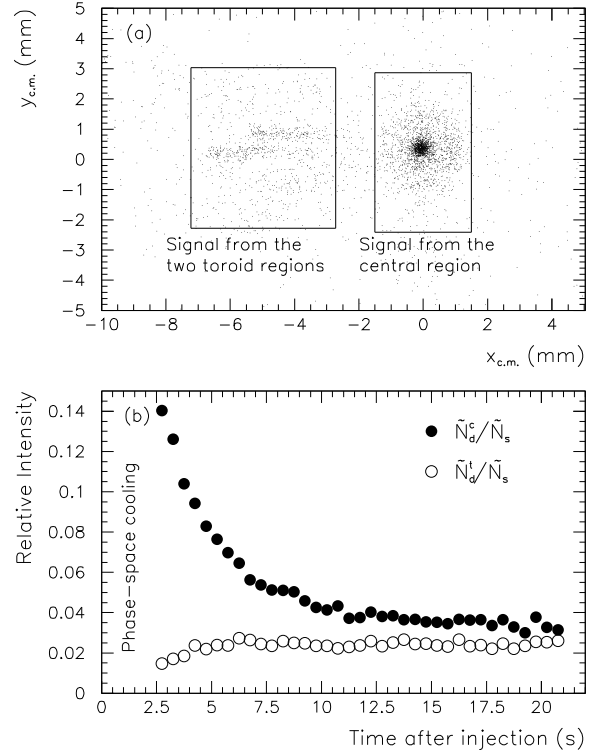}
\caption{
  DR (two-body) events recorded with the
  fragment imaging technique.  (a) Two-dimensional distribution of
  events according to their c.m.\ positions $x_{\rm c.m.}$, $y_{\rm
    c.m.}$.  (b) Time evolution of the signal from the central part
  (filled circles) and the toroid part (open circles) of the electron
  cooler with normalization to the apparent one-body events
  ($\tilde{N}_s$).}
\label{cm}
\end{figure}

Assuming that the fragment masses are identified, the coordinates of
the center of mass (c.m.)\ can be determined for correlated two-body
events.  The c.m.\ coordinates represent the transverse motion of the
interacting $^3$He$^4$He$^+$ ions, propagated on a straight line from
the location of the DR event to the detector.  Although the present 
fragment imaging system does not allow the masses to be identified,
the c.m.\ coordinates can be reconstructed for, on the average, 50\%
of all events by assigning the masses of 3 and 4 amu, respectively,
arbitrarily to the individual fragments in each observed correlated
pair.  The observed pattern of c.m.\ positions in the transverse
plane, shown in Fig.\ \ref{cm}(a) for an electron energy $E_d=0$,
reveals different contributions to the low-energy DR rate.  One
contribution has circular symmetry and is located near the nominal
beam axis of the straight storage ring section to which the detector
is aligned (zero point of the $x$-$y$ plane).  Two additional,
horizontally elongated structures appear at $-7~{\rm mm} < x_{\rm
  c.m.} < -3$ mm and $-0.5~{\rm mm} < y_{\rm c.m.} < 1.5$ mm.  While
all contributions must arise from electron interaction (see Sec.\ 
\ref{sec:drtime}), the circular part can be assigned to the straight
(central) overlap section of the electron cooler, while the two
elongated structures clearly originate from the two toroid regions,
where the electron beam is guided in and out of the ion orbit.  In
fact, the ion beam is horizontally deflected by the magnetic field of
the toroids \cite{lampert1996}, whose bending plane is oriented
vertically.  As expected, the main contribution from the toroids does
not arise close to the beam merging region, where the electron
interaction energy is still close to $E_d=0$, but further up or down
the bends, where the interaction energy reaches values close to 7 eV,
i.e., around the peak in the DR rate coefficient 
(see Fig.\ \ref{fig:raw}(a)); here, the ion
beam is already bent substantially from its aligned direction in the
overlap region.  Note that in Fig.\ \ref{cm}(a) the narrow circular
peak near the zero point of the $x$-$y$ plane is surrounded by a more
diffuse distribution of events which represents those events for which
the arbitary assignment of the individual fragment masses was
incorrect.  Yet, by suitable cuts on the c.m.\ positions as indicated
in Fig.\ \ref{cm}(a), the fragment imaging data for $E_d=0$ can be
decomposed into separate contributions from the linear overlap region
of the electron cooler and from the toroid regions.

The described assignment of the c.m.\ position ranges is further
supported by the time depencence of the rates $\tilde{N}_d^c$ and
$\tilde{N}_d^t$ for the central and the toroid region, respectively.
These rates are shown with a normalization to the observed rate of
single events, $\tilde{N}_s$, in Fig.\ \ref{cm}(b).  The relative rate
$\tilde{N}_d^c/\tilde{N}_s$ shows a significant time variation over up
to 20 s, similar to the variation of the DR rate constant at $E_d=0$
discussed in Sec.\ \ref{sec:drtime}.  In contrast,
$\tilde{N}_d^t/\tilde{N}_s$ settles to a constant value much faster,
in consistency with the observation of a fixed DR rate constant for
$E_d=7.3$ eV in Sec.\ \ref{sec:drtime}.

\subsubsection{Fragment imaging distributions}
\label{sec:im-dist}

The analytical description of transverse (projected) interparticle
distance distributions for DR fragment imaging of diatomic molecular
ions has been discussed in detail earlier \cite{amitay1999}.  At
vanishing electron collision energy ($E_d=0$; thermal electron energy
neglected) the kinetic energy release $E_k$ for well defined initial
state $v,J$ and final atomic states $n,n'$ can be written as
\begin{equation}
E_{k}^{vJnn'} = E_{vJ}-E_{nn'},
\label{Ek}
\end{equation}
where $E_{vJ}$ denotes the initial molecular energy level and
$E_{nn'}$ the asymptotic energy level of the atomic fragments.  Each
contribution with a given kinetic energy release is associated with a
characteristic distribution $P_{vJnn'}(D)$ \cite{amitay1999} of
projected distances, displaying an endpoint at the maximal transverse
distance
\begin{equation}
D_{\rm max}  = s_{\rm max} \frac{M(^3{\rm He})+M(^4{\rm He})}
{\sqrt{M(^3{\rm He})M(^4{\rm He})}}
\sqrt{\frac{E_{k}^{vJnn'}}{E_i}},
\end{equation}
where 
$s_{\rm max}$ denotes the maximum possible distance from a DR
interaction point to the detector (in the present setup $s_{\rm
  max}=7.17$ m, corresponding to the far end of the central electron
interaction region as seen from the fragment imaging detector).  With
given molecular level populations $p_{vJ}(t)$ and with the DR rate
coefficients $\alpha_{\rm DR}^{vJnn'}(0)$ for specific initial and
final states, the total spectrum of measured projected distances can
be written as the superposition
\begin{equation}
 P(D,t) = K(t) \sum_{vJnn'} p_{vJ}(t) 
    \alpha_{\rm DR}^{vJnn'}(0) P_{vJnn'}(D), 
\label{dist_allJ}
\end{equation}
where $K(t)$ is a normalization factor.  As contributions from various
rotational levels cannot be resolved, the distinct features in the
projected distance distribution are associated with individual
vibrational levels and individual atomic quantum numbers $n,n'$.
Using $J$-averaged DR rate coefficients $\tilde{\alpha}_{\rm
  DR}^{vnn'}(0)$ and distance distributions $\tilde{P}_{vnn'}^{(T_{\rm
    rot})}(D)$, taking into account the rotational energy and thermal
rotational level populations for a suitable temperature $T_{\rm rot}$
\cite{amitay1999}, the superposition of Eq.\ (\ref{dist_allJ}) is then
represented as
\begin{equation}
 P(D,t) = K(t) \sum_{vnn'} p_{v}(t) 
    \tilde{\alpha}_{\rm DR}^{vnn'}(0) \tilde{P}_{vnn'}^{(T_{\rm
    rot})}(D).
\label{dist_all}
\end{equation}
The distributions of relative distances between fragments from DR
reactions as obtained with the fragment imaging technique reveal
information on the relative contributions to the overall DR rate from
different reaction channels $vnn'$, as characterized by
$p_{v}(t)\tilde{\alpha}_{\rm DR}^{vnn'}(0)$.  For assigned vibrational
levels $v$, these data also yield the branching ratios to the possible
atomic final states.

\begin{figure}[htbp]
\centering
\includegraphics[width=3.0in]{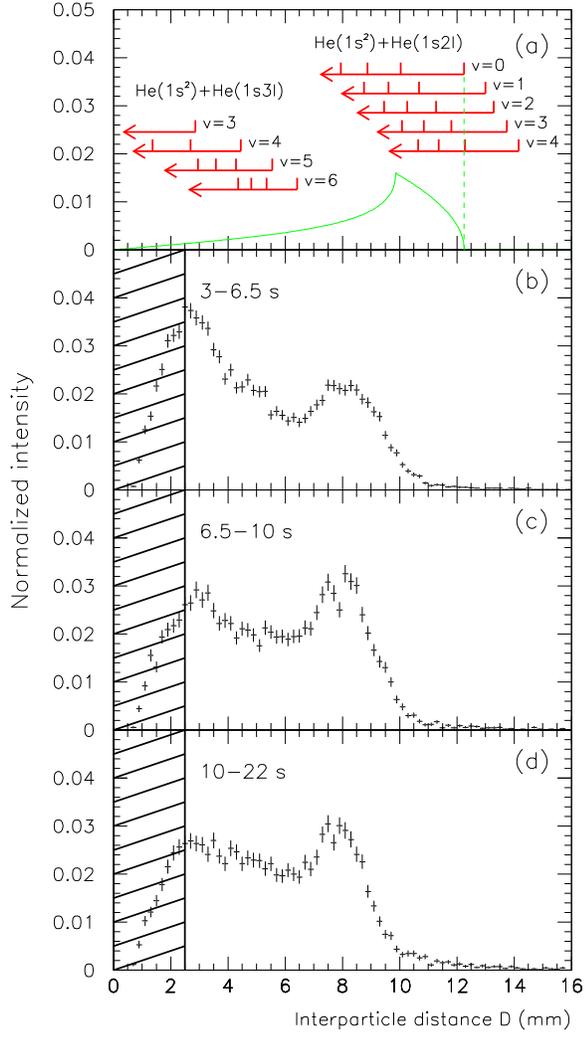}
\caption{
  Distributions of projected interparticle distances ($D$) from DR
  (two-body) events in the central part of the electron cooler at zero
  detuning energy $E_d$.  (a) Analytical form of the distribution for
  a special case (see text) and end points (marked by vertical ticks on 
  arrows) for various possible initial-to-final-state channels.  
  The lower frames show experimental
  results for time intervals of (b) 3--6.5 s (c) 6.5--10 s, and (d)
  10--22 s.  The hatched area at low distance marks the region of
  limited detection due to overlapping light spots on the phosphor
  screen.  Ion energy $E_i=3.36$ MeV; electron density
  $n_e=5.5\times10^6$ cm$^{-3}$.}
\label{distance_fig1}
\end{figure}

The DR fragment imaging measurements discussed below were performed at
an ion energy of $E_i=3.36$ MeV using a detuning energy $E_d=0$ (where
the electron density amounted to $n_e=5.5\times10^6$ cm$^{-3}$).  
Figure \ref{distance_fig1}(a) shows as an example the analytical shape
of the projected distance distribution for $v=0$, $J=0$ and for the
final channel He($1s^2$) + He($1s2s ^3S$) ($E_k= 2.41$ eV), together
with the end points for other possible final states with $n=2$ and 3,
as reached from the vibrational levels $v=0$--6.  As seen in Fig.\ 
\ref{he_curves}, the zero-energy DR of $^3$He$^4$He$^+$ ions in the
vibrational states $v\ge3$ can also lead to the $n=3$ final atomic
states, associated with much smaller kinetic energy release.  In the
observed projected distance distributions, contributions with end
points at small $D$ therefore indicate the presence of vibrationally
excited ions in levels $v\ge3$, and can {\em only} arise from such
levels.

The projected distance distributions observed with a stored
$^3$He$^4$He$^+$ beam for different time intervals after the injection
are shown in Fig.\ \ref{distance_fig1}(b)--(d).  A prominent
contribution at low distances $D$, clearly arising from vibrationally
excited ions in $v\ge3$, can be seen at all times, although its
relative size decreases substantially during the first $\sim$10 s of
storage time.  The peak at large distances becomes narrower at later
storage times.  Two regions showing different types of temporal
behavior can be identified, with a border at $D\sim7$ mm.  Limited
detection of fragment pairs, as discussed in Sec.\ 
\ref{sec:imaging-conditions}, occurs at $D<2.5$ mm.

\begin{figure}[htbp]
\centering
\includegraphics[width=3.0in]{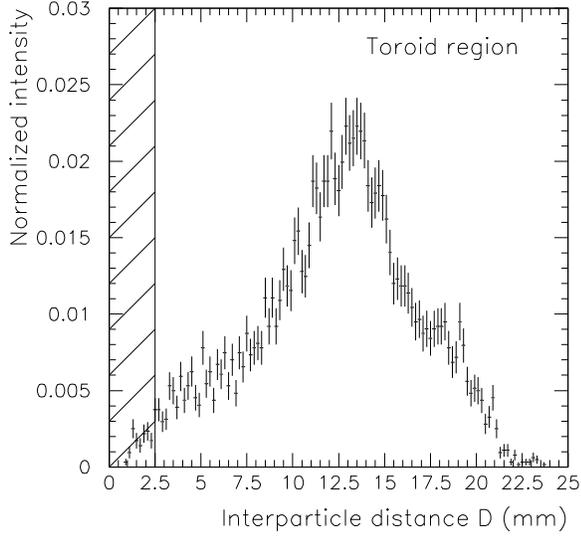}
\caption{
  Fragment imaging spectrum for DR
  from the toroid regions of the electron cooler for the same
  conditions as in Fig.\ \ref{distance_fig1} and for storage times of
   10--22 s. }
\label{rdist_ex}
\end{figure}

The fragment distributions of Fig.\ \ref{distance_fig1} include data
only from the central part of the electron--ion interaction region,
corresponding to the ``central region'' c.m.\ cut shown in Fig.\ 
\ref{cm}(a).  For the toroid regions, much different projected
distance distributions are observed. As seen in Fig.\ \ref{rdist_ex} 
they are characterized by much higher energy release, 
dominated by the dissociation dynamics on the high-energy peak 
of the DR rate coefficient between 5 and 9 eV.

It is interesting to note that the projected distance distribution
from the toroid regions in Fig.\ \ref{rdist_ex} appears to be only
slighty affected by the limited detection of fragment pairs with small
$D$.  This opens up the opportunity of finding the fraction of events
suppressed by limited detection at small $D$ in the fragment
distributions from the central part of the interaction region.  
Using the solid state detector, offering unit detection efficiency, the size
of the toroid contribution relative to the DR rate from the central region of
the electron cooler was determined in Sec.\ \ref{sec:drenergy} to
amount to 
$\tilde{\alpha}_{\rm DR}^{{\rm tor}}(0)/\tilde{\alpha}_{\rm DR}(0)=0.34(6)$.  
On the other hand,
Fig.\ \ref{cm}(b) shows a ratio between the double-hit events from the
toroid and the central region of
$\tilde{N}_d^t/\tilde{N}_d^c=0.67(10)$ for $t>15$ s.  Assuming that
the toroid contribution on the imaging detector, $\tilde{N}_d^t$, is
not influenced by limited detection, while the central contribution is
reduced through this effect by a factor of $1-\tilde{f}_d$, the
missing fraction $\tilde{f}_d$ can be determined to
\begin{equation}
\tilde{f}_d = 1 - \frac{\tilde{\alpha}_{\rm DR}^{{\rm tor}}(0)/\tilde{\alpha}_{\rm
  DR}(0)}{\tilde{N}_d^t/\tilde{N}_d^c}
= 0.49(11)
\label{eq:f_tilde}
\end{equation}
for times larger than 15 s after injection.  In Sec.\ 
\ref{sec:absrate} this number will be used to evaluate the absolute
DR rate coefficient for the $v=0$ level.

\section{Rovibrational excitation and cooling of the stored ion beam}
\label{sec:rovib_clarify}

As illustrated in Sec.\ \ref{sec:beam_evolution_rad}, radiative
thermalization with the blackbody radiation at 300 K would make the
ion beam dominated by the vibrational ground state with
$p_{v=0}>99.9$\% after 10 s of storage, leaving $<$0.1\% in all higher
vibrational states [see Fig.\ \ref{radiative1}(a)].  However, the
rotational radiative relaxation is expected to be much slower [Fig.\ 
\ref{radiative1}(b)].  Moreover, a number of processes beyond
radiative transitions were discussed in Sec.\ \ref{sec:beam_evolution_coll} 
that may modify the rovibrational populations.  Hence, internal excitations
in the stored molecular ion beam deserve careful consideration.

The Coulomb explosion measurements described in Sec.\ 
\ref{sec:exp_CEI} confirm the vibrational relaxation of the ions with
a possible remaining excited-state population at the percent level for
the case that no electrons are present. In the presence of electrons a
similar vibrational relaxation is seen, but the observational
uncertainties would allow the excited-state populations to exceed
the 10\% level.  Based on these measurements and considering the
expected large increase of the He$_2^+$ DR rate coefficient for
increasing vibrational excitation, the excited state populations in
the stored ion beam can still be sufficiently large to cause
significant contributions to the observed neutral fragment rates.
Regarding rotations, the Coulomb explosion imaging method is
unfortunately not suited to characterize the excited level
populations.

In this section, a series of arguments and dedicated measurements is
described which attempts to clarify in detail the influence of
internal molecular excitations on the present data and to identify
processes beyond blackbody-driven radiative interactions that
determine the evolution of rovibrational populations in the stored ion
beam.  From the evidence obtained, it then becomes possible to extract
quantitative, initial- and final-state specific information on the DR
of ${^3}$He${^4}$He$^+$ (Sec.\ \ref{resultsB}).  A further discussion
using a model description of ro-vibrational excitation and
de-excitation (Sec.\ \ref{sec:resultsC}) will finally allow the cross
sections of the underlying inelastic processes to be quantitatively
characterized.

\subsection{Time dependence of zero energy DR rate}
\label{sec:total-DR-rate}               

\begin{figure}[htbp]
\centering
\includegraphics[width=3.0in]{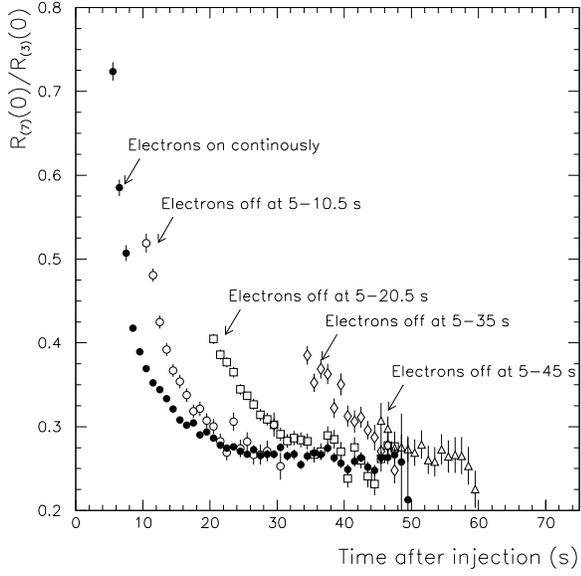}
\caption{
  Measured DR rate $\rdr(0)$ normalized to $\rdei(0)$ for situations
  where the electron beam was continuously on (filled circles), and
  switched off at 5--10.5 s (open circles), 5--20.5 s (open squares),
  5--35 s (open diamonds), and 5--45 s (open triangles).  Ion energy
  $E_i=7.28$ MeV; electron density $n_e=1.2\times10^7$ cm$^{-3}$.}
\label{cooling_fig1}
\end{figure}

Figure \ref{cooling_fig1} shows the ratio $\rdr/\rdei$ at $E_d=0$ for
conditions where the electron beam was either continuously on or
switched off for different segments of time after the end of the
phase-space cooling.  Following Eqs.\ (\ref{eq:master2}) and
(\ref{eq:master3}) and noting the time independence of $\tilde{k}_{\rm
  DE}(0)$ (Sec.\ \ref{sec:dr-rate}), this ratio is proportional to the
rovibrationally averaged DR rate constant $\tilde{k}_{\rm DR}(0;t)$
and hence probes the evolution of the rovibrational population in the
ion beam if the DR rate coefficients of the rovibrational levels
differ from each other [see Eq.\ (\ref{eq:alp_mean_general})].

\begin{figure}[htbp]
\centering
\includegraphics[width=3.0in]{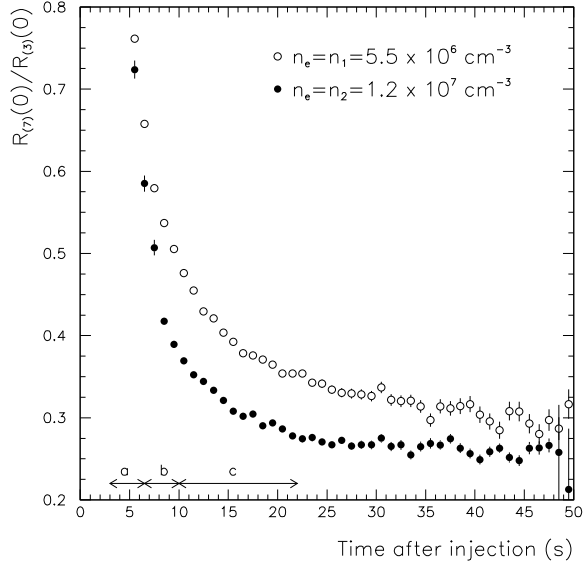}
\caption{Measured DR rates $\rdr(0)$ normalized to $\rdei(0)$ 
  measured with the electron beam continuously on for electron
  densities of $n_e=n_1=5.5\times10^6$ cm$^{-3}$ (open circles) and
  $n_e=n_2=1.2\times10^7$ cm$^{-3}$ (filled circles).  
  To compare the relative rate $\rdr(0)/\rdei(0)$
  at the two different electron densities the data 
  measured for $n_1$ have been scaled with the help of 
  Eq.\ (\ref{eq:master2}), (\ref{eq:master3}), and (\ref{eq:c1})	
  by the factor $n_2(1+1/c_1)/(n_2+n_1/c_1)=1.77$. 	
  Ion energy $E_i=7.28$ MeV.  The time intervals marked a,
  b, and c are those for which fragment imaging spectra are presented
  in Fig.\ \ref{distance_fig1}.}
\label{compare_ne_fig1}
\end{figure}

As seen in Fig.\ \ref{cooling_fig1}, when the electron beam is
continuously on, $\tilde{k}_{\rm DR}(0;t)$ decreases strongly until an
apparently constant level is reached after about 30 s.  The shape of
the curve cannot be well-described by a simple function such as a
single exponential plus a constant or the sum of two exponentials.
When the electron beam has been off for one of the various segments of
time, a larger value of $\rdr/\rdei$ is observed when switching it
back on as compared to the situation where the electron beam is
continuously on.  However, after switching the electron beam on again,
$\rdr/\rdei$ decreases on a time scale of $\sim$10 s to the level
reached with the electron beam continuously on.  The effect of the
electron beam on the rovibrational population is further emphasized by
the data shown in Fig.\ \ref{compare_ne_fig1}, which displays the
value of $\rdr/\rdei$ for two different electron densities.  A higher
electron density leads to a faster approach to the asymptotic DR rate,
while the final asymptotes are similar.  The electron-beam dependent
behaviors illustrated in Fig.\ \ref{cooling_fig1} and
\ref{compare_ne_fig1} clearly show that the DR reaction rate depends
upon the rovibrational state and that the electron beam indeed
influences the rovibrational population in the beam.

Moreover, as seen in Fig.\ \ref{cooling_fig1}, the value of
$\rdr/\rdei$ {\em at the moment} when the electron beam is switched
back on decreases with longer periods without electron cooling.  This proves
unambiguously that the rovibrational population in the ion beam
evolves over tens of seconds even if only radiative transitions and
residual gas collisions can occur.  The electron beam was also
switched off for a similar period as in Fig.\ \ref{cooling_fig1} at
later times after injection (35--50 s), i.e., after $\rdr/\rdei$ had
already reached its asymptotic value.  In this test (not shown in
Fig.\ \ref{cooling_fig1}) no significant change of the ratio
$\rdr/\rdei$ could be seen when switching the electron beam back on.
Hence, on a time scale of 15 s the asymptotic rovibrational population
reached after continuous electron interaction is not altered under the
combined influence of blackbody radiation and residual gas collisions.


Using the estimates on the rovibrational evolution given in Sec.\ 
\ref{sec:beam_evolution_rad} it is possible to understand more precisely
the origin of the signal evolutions shown in Fig.\ \ref{cooling_fig1}
and Fig.\ \ref{compare_ne_fig1}.  Radiative transitions drive the
vibrational population of the stored ions to stabilize within a few
seconds after injection, and the fast decay of the DR rate up to
$\sim$10 s could well represent the vibrational stabilization.  That
the vibrational stabilization is observed for longer times than seen
with the CEI technique is expected since the DR measurements are much
more sensitive to higher vibrational states, given the larger DR cross
sections for these as compared to the vibrational ground state $v=0$
\cite{carata1999}.  The final vibrational populations in the beam
cannot be estimated based on the results in Fig.\ \ref{cooling_fig1}
and Fig.\ \ref{compare_ne_fig1}, but will be discussed in Sec.\ 
\ref{sec:initial-to-final}.

Based on its timescale, the following evolution of the DR signal over
tens of seconds, i.e., beyond 10 s in Fig.\ \ref{cooling_fig1} and
Fig.\ \ref{compare_ne_fig1}, must originate from changes in the {\em
  rotational} populations.  As a first consequence, this demonstrates
that the DR cross section depends on the initial rotational state of
the interacting ion.  In Sec.\ \ref{sec:beam_evolution_coll} two possible
effects of the electron beam on the rotational populations were
discussed: preferential depletion of specific $J$ levels by the 
DR reaction, or their electron-impact excitation and de-excitation [Eq.\ 
(\ref{eq:general_ex})].  

Preferential depletion by the DR reaction would produce a significant
effect on the beam decay, which was not observed in the measurements of 
Sec.\ \ref{sec:dr-rate} (see Fig.\ \ref{electron_off}).
Moreover,
the electron induced rate variation seen in Fig.\ \ref{cooling_fig1}
when switching the electron beam back on occurs over a time scale of
$\sim$10 s.  With the applied electron density of $n_e=1.2 \times
10^{7}$ cm$^{-3}$, the rate coefficients involved in the underlying
changes of the rotational level populations should have a size of
$\sim$(10 s\,$\cdot \eta n_e)^{-1} = 3\times10^{-7}$ cm$^{3}$/s.  On
the other hand, the observed absolute DR rate coefficient at $E_d=0$
(including contributions from excited states with $v\ge3$) amounts to
only $\sim${}$3\times10^{-9}$ cm$^{3}$/s [see Fig.\ 
\ref{rate_dr_de}(a)], much smaller than required to depopulate a
significant part of the ion beam.

These arguments leave electron-induced cooling [Eq.\ 
(\ref{eq:general_ex})] as the mechanism driving the evolution of the
rotational states in the stored ion beam.  It should be noted that the
time scale for variations of the DR rate coefficient through the
radiation field alone, as revealed by the starting points of the
various cooling curves in Fig.\ \ref{cooling_fig1}, approaches several
tens of seconds at later times after injection and probably reaches
$\gtrsim$100 s as revealed by the additional test described above,
when electrons were switched off at times of 35--50 s.  In connection
with the radiative lifetimes shown in Fig.\ \ref{lifetime}, this
indicates a rotational distribution dominantly composed of levels with
$J\lesssim6$, i.e., rotational temperatures of $\lesssim$300 K.

\subsection{Assignment and time dependence of DR fragment imaging spectra}
\label{sec:initial-to-final}

\begin{figure}[htbp]
\centering
\includegraphics[width=3.0in]{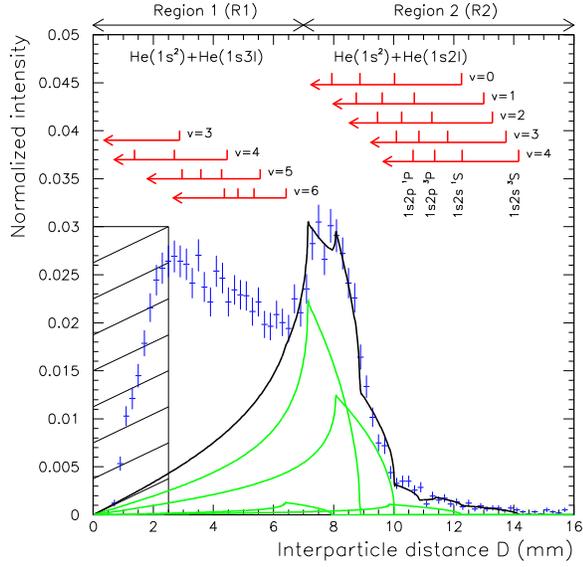}
\caption{
  Assignment of DR reaction channels to the data of Fig.\ 
  \ref{distance_fig1}(d), taken at 10--22 s after injection.  The
  vertical lines grouped for different intitial vibrational levels $v$
  show the endpoints of the distributions for various DR reaction
  channels as explained in the text.  The black curve shows a
  least-squares fit to the data, explained in Sec.\ 
  \ref{initial_final}; the gray lines show the contributions of the
  individual DR channels from $v=0$ included in this fit.  The hatched
  area at low distance marks the region of limited detection due to
  overlapping light spots on the phosphor screen.  Ion energy
  $E_i=3.36$ MeV; electron density $n_e=5.5\times10^6$ cm$^{-3}$.}
\label{distance_fig3}
\end{figure}

The DR fragment imaging spectra shown in Fig.\ \ref{distance_fig1}
were taken under similar cooling conditions as the curve shown by open
symbols in Fig.\ \ref{compare_ne_fig1}, using the same electron
density ($5.5\times10^6$ cm$^{-3}$) although the ion beam energy was
reduced in order to enhance the fragment separation in the detector
plane.  The time slices of the three imaging spectra are marked by
arrows in Fig.\ \ref{compare_ne_fig1}.  In the following, we will
first give a preliminary assignment of the structures in the imaging
distribution from the latest time slice, enlarged in Fig.\ 
\ref{distance_fig3}; then we will consider the time evolution of the
individual signals in two regions of this distribution with a
subdivision at the interparticle distance $D=7$ mm.

As stated earlier, the broad structure at $D<7$ mm in the projected
distance distribution, marked as Region 1 in Fig.\ 
\ref{distance_fig3}, can only arise from the DR of ions initially in
vibrational states with $v \ge 3$, proceeding to products where one He
atom emerges in an excited state with a principal quantum number $n
\ge 3$.  Vertical lines, grouped according to the initial levels with
$v\geq3$ in this region, indicate the end points of the expected shapes
$P_{v0nn'}(D)$ [cf.\ Eq.\ (\ref{dist_allJ})] for transitions from a
given vibrational level to the different fine structure terms of the
He$(1s^2\,{^1}S)$ + He$(1s3l)$ final channel.  The assignment of the
signal in Region 2 ($7\,{\rm mm} < D < 16\,{\rm mm}$) is at this point
ambiguous.  The peak at 7--12 mm can be well described by reactions
from the ion's vibrational ground state ($v=0$) to final states with
an excited He atom in $1s2s\,{^3}S$, $1s2s\,{^1}S$, $1s2p\,{^3}P$ and
$1s2p\,{^1}P$. However, the peak structure can also be described by
including substantial contributions from $v=1$.  Moreover, to fully
describe the tail to higher distances, some contributions from
vibrational states $v \ge 3$ are certainly needed, as indicated in the
figure.  In Sec.\ \ref{sec:pump-probe} below it will be argued that
the major part of the signal in Region 2 indeed must arise from
transitions from $v=0$.

Figure \ref{dist_vs_time_fig1}(a) shows the time evolution of the
integrated signals in Region 1 and Region 2, $\tilde{N}_d^{\rm R1}$
and $\tilde{N}_d^{\rm R2}$ after normalization to the detected number
of one-body events, $\tilde{N}_s$.  The signals from the two regions
have very different time dependencies in the first $\sim$6.5 s.  While
the signal from Region 1 decreases strongly, the signal from Region 2
shows a slight increase.  From 6.5 s onwards both signals decrease,
the signal from Region 1 essentially reaching a constant level after
$\sim$12 s, while that from Region 2 continues to decrease slowly over
the full observation period up to 22 s.  This tendency is emphasized
in Fig.\ \ref{dist_vs_time_fig1}(b) where the ratio of the two
signals, $\tilde{N}_d^{\rm R1}/\tilde{N}_d^{\rm R2}$, is displayed as
a function of time.  Thus, the increase in the ratio at the late times
demonstrates the different time evolution at times beyond $>$10 s.

The fact that the signals in Region 1 and 2 have very different time
structures for the first 6.5 s proves that they originate from
different parts of the rovibrational population in the ion beam at
these times and indicates that the signal in Region 2 arises from the
lower vibrational levels, possibly up to $v=2$.  This is also
consistent with the fast vibrational stabilization, taking at most 12
s, imposed by the radiative decay of excited vibrational states as
discussed in Sec.\ \ref{sec:beam_evolution_rad}.  The time evolution on
the longer time scale ($\ge$10 s) can be unambiguously assigned to
rotational states with different DR cross sections.  Since the signal
from Region 2 shows a time dependence after 10 s, while an apparent
stabilization is seen for the signal in Region 1 ($v\ge3$), the
rotational dependence of the DR cross section seems to be more
important for the vibrational ground state (possibly including $v=1$
and 2 which have not yet been excluded at this point) than for the
higher vibrational states.

The distribution of interparticle distances in Fig.\ 
\ref{distance_fig3} shows that vibrational levels with $v \ge 3$
(Region 1) contribute significantly to the DR signal even after $>$10
s of storage time, in contrast to the expected effect of radiative
relaxation (Sec.\ \ref{sec:beam_evolution_rad}), which would yield
insignificant excited state populations not contributing to the DR
signal despite their predicted large cross sections.  In consequence,
since the populations cannot be determined by radiative interactions
alone, the stored ions must experience vibrational excitation by some
further mechanism(s).

\subsection{Fragment imaging pump-probe experiments}
\label{sec:pump-probe}

\begin{figure}[htbp]
\centering
\includegraphics[width=3.0in]{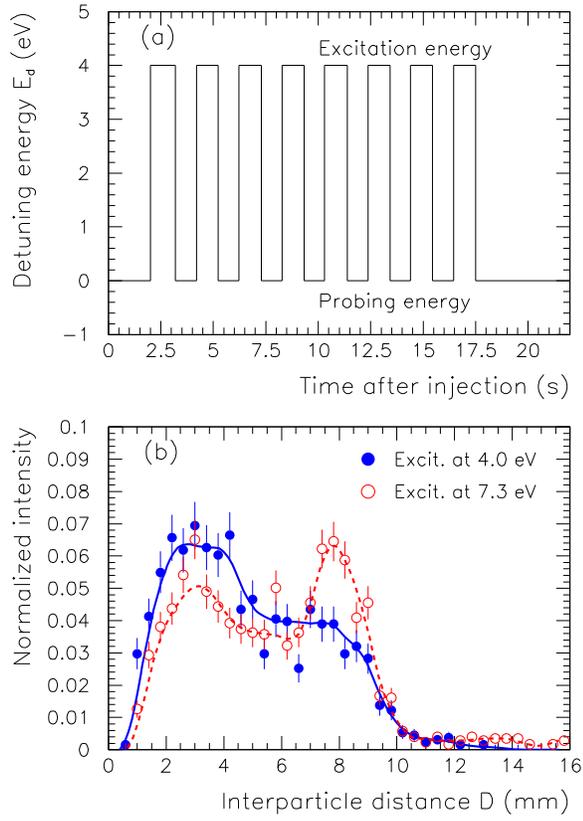}
\caption{
  Pump-probe experiments using the fragment imaging method.  (a)
  Schematic illustration of the operation of the electron cooler.  (b)
  Projected distance spectrum obtained in the 4 probing periods within
  the time interval 9--16 s, each time following excitation with
  electrons at 4.0 eV (filled circles) and 7.3 eV (open circles).  
  The shapes of the two distributions are emphasized 
  (to guide the eye) by the solid and the dotted curves. 
  Ion energy $E_i=3.36$ MeV; electron density $n_e=5.5\times10^6$
  cm$^{-3}$.}
\label{pump-probe}
\end{figure}

The observations that the signal in Region 1 arises from higher
vibrational states ($v \ge 3$) and that the major part of Region 2
originates from DR of the lower vibrational states makes it possible
to use a pump-probe type experiment to investigate the effect of the
electron beam and of residual-gas collisions on the vibrational
population.  The operation of the electron cooler during these
measurements is schematically shown in Fig.\ \ref{pump-probe}(a).
Following each injection, the energy of the electron beam is altered
in a regular time pattern, spending several time intervals of 1 s each
at an excitation energy [$E_d=4.0$ eV in Fig.\ \ref{pump-probe}(a)]
and interrupting these excitation periods by probing intervals of 1 s
each, where projected fragment distance distributions are recorded at
$E_d=0$ (probing energy).  The effect of electron excitation at
different energies is illustrated in Fig.\ \ref{pump-probe}(b) by
showing the distribution of projected distances accumulated in the
last four probing intervals [$t=9$--16 s, see Fig.\ 
\ref{pump-probe}(a)] at two different excitation energies.  The ratio
of events in the Regions 1 and 2 is clearly different for the two
excitation energies, Region 1 (originating from $v\ge3$) dominating
the spectrum much more for 4.0 eV excitation than for 7.3 eV
excitation.  

\begin{figure}[htbp]
\centering
\includegraphics[width=3.0in]{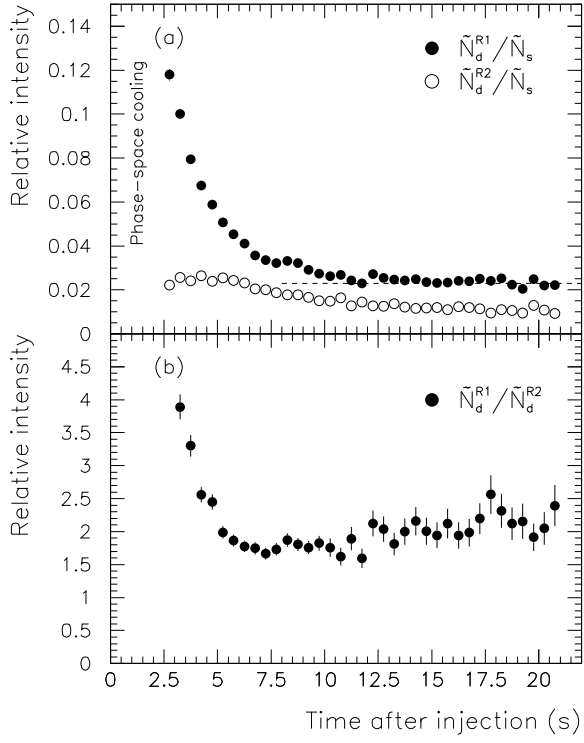}
\caption{
  Time evolution of the dissociative recombination signal at $E_d=0$
  as observed with the fragment imaging technique (contribution from
  the central part of the electron cooler only).  (a) Intensities of
  the signals from the Regions 1 and 2 of the fragment imaging
  distribution (cf.\ Fig.\ \ref{distance_fig3}) relative to the
  observed number $\tilde{N}_s$ of one-body imaging events.  (b) Ratio
  of the signal intensities from Regions 1 and 2.}
\label{dist_vs_time_fig1}
\end{figure}

In the following, we will quantify the DR rates associated with
Regions 1 and 2 by the integrated counts of two-body events in the
respective regions, $\tilde{N}^{\rm R1}_d$ and $\tilde{N}^{\rm R2}_d$,
each normalized to the number $\tilde{N}_s$ of single-hit events
during the same counting period.  Obviously, the signals in Regions 1
and 2 are not completely independent since the reactions contributing to
the signal in Region 2 partly give intensity to Region 1.  However, it
is evident from Fig.\ \ref{dist_vs_time_fig1} that the signal in
Region 2 evolves slowly with time, and to a first approximation its
contribution to Region 1 can be considered constant.  With this
observation, the relative intensity $\tilde{N}^{\rm R1}_d/\tilde{N}_s$
for Region 1 yields a useful probe signal for the vibrational
population in $v\ge3$, provided one keeps in mind that it contains an
offset contribution proportional to the corresponding signal for
Region 2.  It should also be stressed that, compared to the continuous
operation used in Secs.\ \ref{sec:total-DR-rate} and
\ref{sec:initial-to-final}, the switching operation in the pump-probe
measurements leads to somewhat different average effects of the
electron beam on the rovibrational populations in the ion beam, so
that the rates $\tilde{N}^{\rm R1,R2}_d/\tilde{N}_s$ cannot be
directly compared to the corresponding quantities of the previous
measurements at the same storage times.

\begin{figure}[htbp]
\centering
\includegraphics[width=3.0in]{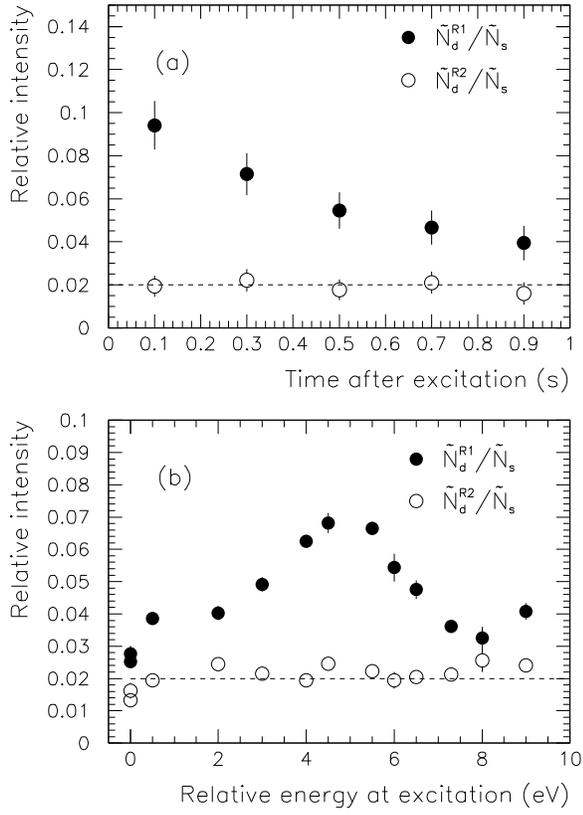}
\caption{
  Electron induced vibrational excitation as studied with the DR fragment
  imaging pump-probe experiments.  The DR signals occurring in Region
  1 (filled circles) and Region 2 (open circles) (cf.\ Fig.\ 
  \ref{distance_fig3}) are shown (a) as a function of time after
  excitation with electrons at 4 eV, and (b) as a function of the
  excitation energy.  The counts were collected from the 4 probing
  periods for $\ge 9$ s after injection; the counts in Regions 1 and 2
  have been normalized to the observed number $\tilde{N}_s$ of
  one-body imaging events.}
\label{pump_probe_electron}
\end{figure}

The association of Region 1 with vibrationally excited ions is
supported by observing the time dependence of the signal
$\tilde{N}^{\rm R1}_d/\tilde{N}_s$ following the end of an excitation
period.  For this purpose, the data collected in the probing intervals
are subdivided into time bins counting from the start of the probing
period.  The temporal variation of the signals from both Regions after
the end of excitation is shown for an excitation energy of 4 eV in
Fig.\ \ref{pump_probe_electron}(a).  The DR signal from Region 1 is
enhanced right after the electron excitation period and then shows a
significant, rapid decrease.  The time constant for the decay of this
signal (containing a constant offset due to the contributions from
Region 2, as discussed above) is consistent with the expected
radiative lifetime of the levels $v=3$--5, lying in the range of
0.4--0.25 s as shown in Fig.\ \ref{lifetime}.  Remarkably, no
significant time variation (above a 10\% level) is seen on the DR
signal from Region 2.  This provides a strong experimental evidence
that the major part of this signal originates from the vibrational
ground state ($v=0$).  In fact, the radiative lifetimes of all excited
$v$ levels (populated either directly or by the radiative cascade from
$v\ge3$) are short enough that significant contributions to the
Region-2 DR signal from such levels should reveal themselves by a
decaying component in the time-dependent pump-probe measurement of
Fig.\ \ref{pump_probe_electron}(a).  It should also be noted that the
constant value of $\tilde{N}^{\rm R2}_d/\tilde{N}_s$ [marked by a
dashed line in Fig.\ \ref{pump_probe_electron}(a)] is slightly higher
than obtained at the same storage times when the electron beam at zero
energy was applied continuously [Fig.\ \ref{dist_vs_time_fig1}(a)].
This is consistent with the interpretation of a strong electron
induced rotational de-excitation at zero energy, as discussed in Sec.\ 
\ref{sec:total-DR-rate}.

The variation of the vibrational excitation with the electron energy is
displayed in Fig.\ \ref{pump_probe_electron}(b).  The signal from
Region 1 is clearly energy dependent with a peak at $\sim$5 eV.
The mechanism underlying this observed electron-induced vibrational
excitation will be discussed in Sec.\ \ref{sec:res_de_exc}.  The
signal originating from Region 2 has no energy dependence, confirming
again that this part of the DR signal originates predominantly from
$^3$He$^4$He$^+$($v=0$) ions.  Note that the energy range scanned in
this measurement covers an essential part of the electron energies
present in the toroid regions of the electron cooler when the detuning
energy in the central part of the electron cooler is close to zero.

\begin{figure}[htbp]
\centering
\includegraphics[width=3.0in]{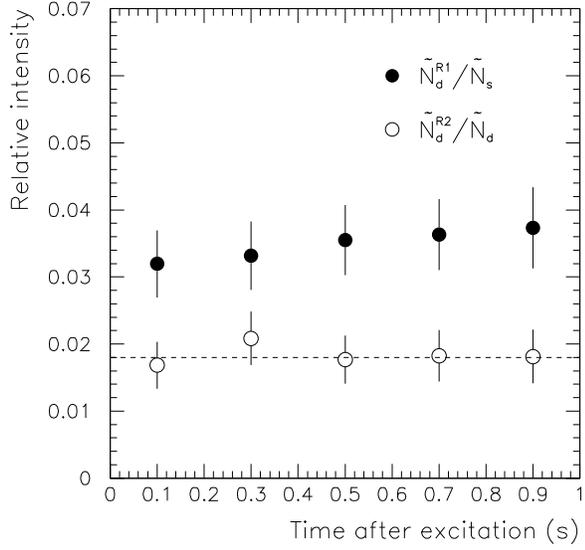}
\caption{
  Residual-gas induced vibrational excitation as studied with the
  fragment imaging pump-probe experiments.  The DR signals occurring
  in Region 1 (filled circles) and Region 2 (open circles) (cf.\ Fig.\ 
  \ref{distance_fig3}) are shown as a function of time after turning
  on the electron beam again, following an off-period of 1 s. The
  counts were collected and normalized as in Fig.\ 
  \ref{pump_probe_electron}.}
\label{pump_probe_gas}
\end{figure}

Fragment imaging measurements with a similar timing cycle were
performed to probe the effect of leaving the $^3$He$^4$He$^+$ ions
under the influence of residual gas collisions only.  To this end,
following each injection, the probing periods were again arranged as
in Fig.\ \ref{pump-probe}(a), but instead of stepping up the electron
energy the electron beam was switched off for 1 s between the probing
periods.  Figure \ref{pump_probe_gas} shows that under these
conditions the probe signal from Region 1 is essentially constant,
possibly with a slight temporal increase showing the effect of
vibrational excitation from the toroid regions after the electrons
have been switched back on.  The signal ratio between Regions 1 and 2
during the probing period continuously stays at $\sim$2, a value close
to that found when the electron beam was continuously on [$t>9$ s in
Fig.\ \ref{dist_vs_time_fig1}(b)].  Altogether, the dominant part of
the observed vibrational excitation persists in the stored
$^3$He$^4$He$^+$ beam even during the absence of the electron beam
(over times that exceed the radiative lifetime of the excited
vibrational levels).  We conclude that the dominant role in
maintaining a small non-thermal vibrational population in the stored
beam is played by ion collisions with the residual gas.

\section{Results}
\label{resultsB}

\subsection{Low-energy DR: initial-to-final state dynamics}
\label{initial_final}

The absence of a time dependence of 
the rate $\tilde{N}^{\rm R2}_d$ in the pump-probe experiments of Sec.\ 
\ref{sec:pump-probe} shows that at most $\sim$10\% of the signal in
Region 2 ($D>7$ mm) of the DR fragment imaging distributions at
$E_d=0$ can arise from ${^3}$He${^4}$He$^+$ ions in the vibrationally
excited states.  Hence, the shoulders on the right-hand side of the
fragments imaging distribution of Fig.\ \ref{distance_fig3} can be
overwhelmingly assigned to different final states of
${^3}$He${^4}$He$^+(v=0)$ DR.  Consequently, the distribution was
fitted with the individual initial-to-final-state contributions listed
in Table \ref{tableB}, yielding the full line in Fig.\ 
\ref{distance_fig3}.  In particular, the dominant contributions
peaking at $\sim$7.2 mm and $\sim$8.0 mm (cf.\ Fig.\ 
\ref{distance_fig3}) must come from the $^3P$ and $^1S$ states,
respectively.  Contributions to the imaging signal whose shape cannot
be attributed to $v=0$ ions are on the few-percent level; within the
remaining arbitrariness caused by the similarity of the energy release
for the different channels with $v\neq 0$, they were attributed to
certain final states reached from ions with $v=3$ and 4, as indicated
in Table \ref{tableB}.

The fitted signal contributions of Table \ref{tableB} now yield the
branching ratios for DR of ${^3}$He${^4}$He$^+(v=0)$ ions into the
various fine structure terms of the $n=2$ final atomic states.  Table
\ref{table1} lists these ratios, which show a dominance of the
$1s2p\,{^3}P$ and $1s2s\,{^1}S$ fine structure levels; this result
will be further discussed in Sec.\ \ref{sec:disc_branching}.

In addition, it is possible to extract from Table \ref{tableB} the
fraction of all detected two-body events that originates from
${^3}$He${^4}$He$^+$ in the vibrational ground state, which amounts to
$\tilde{I}(v=0)=0.54(5)$.  This result will be used in the next
section to derive the low-energy DR rate coefficient for $v=0$
${^3}$He${^4}$He$^+$ ions.

Finally, two conclusions can be drawn from the low fraction (at most
$\sim$10\%) of signal contributions from vibrationally excited ions in
Region 2.  Firstly, this indicates that the levels $v=1$ and 2, for
which the DR can contribute to Region 2 only ($n=2$ final atomic
states), are weakly populated in the stored ion beam.  In fact, the DR
rate coefficient should grow significantly with increasing $v$ through
the improving Franck-Condon overlap (an increase by at least a factor
of 10 for each quantum of vibrational excitation between 0 and 2 was
predicted in Ref.\ \cite{carata1999}), so that the populations of
excited $v$ levels are likely to be $<$1\%.  Secondly, the levels
$v\geq3$ (likely to have similar populations) do yield a substantial
DR signal in Region 1 ($n\geq3$ final atomic states) but, remarkably,
still seem to contribute very little to Region 2.  Hence, ions in
these levels show a strong preference for DR into the higher-lying
atomic states, avoiding the $n=2$ final levels.

\subsection{Absolute low-energy DR rate coefficient}
\label{sec:absrate}

From the comparison of the ion beam decay rate with and without the
presence of the electron beam, an absolute rate coefficient of
$\tilde{\alpha}_{\rm DR}(E_r)=2.8(4)\times10^{-8}$ cm$^{3}$\,s$^{-1}$
was deduced in Sec.\ \ref{sec:drtime} for the DR on the high-energy
peak at a relative energy of $E_d^r = 7.3$ eV.  The energy-dependent
rate measurements were normalized using this absolute rate and, after
the toroid correction (see Sec.\ \ref{sec:drenergy}), yield for $t>35$
s a zero-energy, rovibrationally averaged DR rate coefficient of
$\tilde{\alpha}_{\rm DR}(0)=2.7(4)\times10^{-9}$ cm$^{3}$\,s$^{-1}$.
Following the assignment and the fit of the DR fragment imaging
distribution as discussed in the previous section, it becomes possible
to determine the fraction of the observed DR rate at $E_d=0$ that
originates from $v=0$ $^{3}$He$^{4}$He$^+$ ions.

The total intensity in the fragment imaging spectrum, corresponding to
the toroid-corrected averaged rate coefficient $\tilde{\alpha}_{\rm
  DR}(0)$, is reduced by the limited detection, as discussed in Sec.\ 
\ref{sec:im-dist}, by a factor of $1-\tilde{f}_d$, where $\tilde{f}_d$
was obtained in Eq.\ (\ref{eq:f_tilde}).  The integrated intensity
$\tilde{I}(v=0)$ of the fitted $v=0$ contributions in the normalized
fragment imaging spectrum can therefore be written as
\begin{equation}
\tilde{I}(v=0) \, = \, 
\frac{p_0\tilde{\alpha}^{v=0}_{\rm DR}(0)}
{(1-\tilde{f}_d)\tilde{\alpha}_{\rm DR}(0)},
\label{eq:I_tilde}
\end{equation}
using the effective contribution $p_0\tilde{\alpha}^{v=0}_{\rm
  DR}(0)$ from $v=0$ ions contained in $\tilde{\alpha}_{\rm DR}(0)$ as 
defined by Eq.\ 
(\ref{eq:alp_mean_general}).
Within the error of the absolute rate coefficient scale of $\sim$20\%
we can here neglect the deviation of $p_0$ from unity.  

Using the values
of $\tilde{I}(v=0)$ and $\tilde{\alpha}_{\rm DR}(0)$ determined above,
we then obtain
\begin{equation} 
\tilde{\alpha}^{v=0}_{\rm DR}(0)=(7.3 \pm 2.1) \times 10^{-10} \,{\rm cm}^3\,{\rm s}^{-1}.
\label{alpha_cor} 
\end{equation}
This rate coefficient is still {\it averaged} over the relaxed
rotational distribution resulting after storage and electron cooling
of 35 s (cf.\ Sec.\ \ref{sec:total-DR-rate}).  The influence of the
rotational relaxation is illustrated by the open symbols in Fig.\ 
\ref{compare_ne_fig1}, keeping in mind that the given result was
obtained with an electron density of $5.5 \times 10^{6}$ cm$^{-3}$.

Assuming a DR cross section varying as $\propto E^{-1}$ up to
$\sim$0.1 eV, the merged-beams rate coefficient
$\tilde{\alpha}^{v=0}_{\rm DR}(0)$ can be converted to a rate
coefficient for a thermal, isotropic electron velocity distribution at
300 K through the division by a factor of $(300$\,K$/T_\perp)^{1/2}
\arctan(T_\perp/T_\parallel-1)^{1/2}/
(1-T_\parallel/T_\perp)^{1/2}=2.2$ that follows from the respective
averages over isotropic and anisotropic Maxwellians.  This yields
$\tilde{\alpha}^{v=0}_{{\rm 300\,K,~DR}} = (3.3 \pm 0.9) \times
10^{-10}$ cm$^3$\,s$^{-1}$.

After subtraction of $\alpha^{v=0}_{\rm DR}(0)$ from the total
observed rate coefficient $\tilde{\alpha}_{\rm DR}(0)$, a difference
of $\Delta\tilde{\alpha}_{\rm DR}(0)=1.9(5)\times10^{-9}$
cm$^{3}$\,s$^{-1}$ remains, that represents the fraction due to
vibrationally excited ions in states $v\geq3$.  The value of
$\Delta\tilde{\alpha}_{\rm DR}(0)$ can be used to estimate the
zero-energy DR rate coefficient for excited vibrational states
$v\geq3$.  Assuming relative populations $p_v=\sum_{J}p_{vJ}<0.01$ for
these levels, we obtain $\tilde{\alpha}^{v\geq3}_{\rm DR}(0) \gtrsim
2\times10^{-7}$ cm$^{3}$\,s$^{-1}$.  

On the other hand, if a value of $\sim$10$^{-6}$ cm$^{3}$\,s$^{-1}$ 
is regarded as an upper limit for the DR rate coefficient of He$_2^+$ 
ions in any vibrational excited level, the relative population of 
levels with $v\geq3$ must be $\geq$0.19 \%.
Hence, it is reasonable to assume excited vibrational
states to occur in the stored ion beam at fractions between $\sim$0.1
and 1\%.

\subsection{Low energy structure of the DR cross section}
\label{sec:low-e-structure}

\begin{figure}[htbp]
\centering
\includegraphics[width=3.0in]{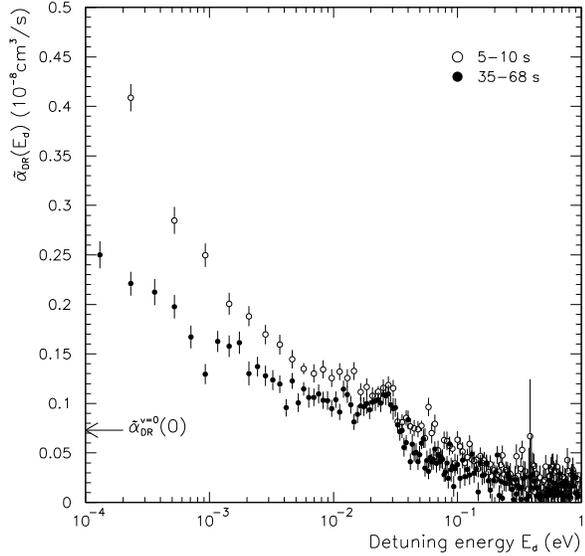}
\caption{ 
  Detailed view of the measured toroid corrected rate coefficients
  $\tilde{\alpha}_{\rm DR}(E_d)$ at low relative energies obtained for
  storage time intervals of 5--10 s (open circles) and 35--68 s
  (filled circles); see Fig.\ \ref{rate_dr_de}(a) for a full view.}
\label{dr_rate_coeff01}
\end{figure}

In the low-energy range the measured rate coefficient
$\tilde{\alpha}_{\rm DR}(E_d)$ shows a strong variation with time as
illustrated for $E_d=0$ in Fig.\ \ref{cooling_fig1}. The measured
energy dependence in two characteristic time windows is shown in Fig.\ 
\ref{dr_rate_coeff01}.  As discussed, contributions from higher
vibrational states are significant even at long storage time, as
illustrated by the indicated size of $\alpha^{v=0}_{\rm DR}(0)$.  The
observed low-energy DR rate coefficient displays a peak-like structure
at $\sim$0.025 eV which persists over all accessible times after
injection, but sharpens at the later times.  The present measurement
shows no indication of further structure in the DR cross section in
the energy range below 1 eV.

\subsection{High energy structure of the DR cross section}
\label{sec:high-e-structure}

\begin{figure}[htbp]
\centering
\includegraphics[width=3.0in]{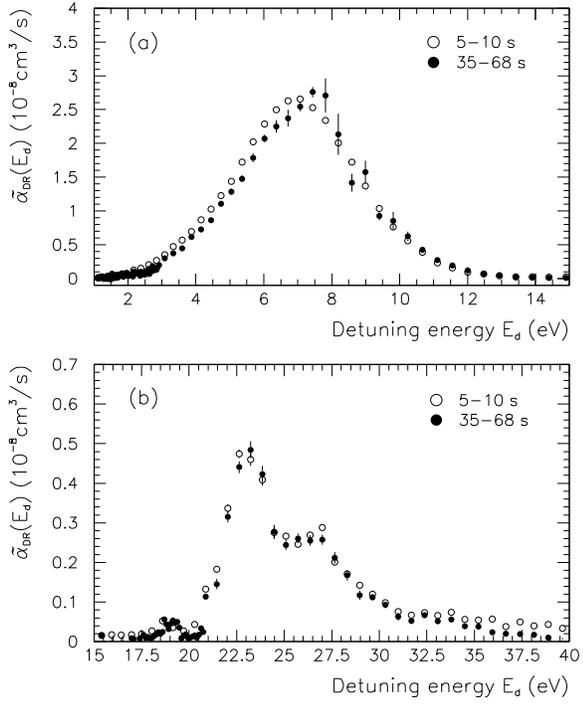}
\caption{
  Detailed view of the measured toroid corrected rate coefficients
  $\tilde{\alpha}_{\rm DR}(E_d)$ obtained for storage time intervals
  of 5--10 s (open circles) and 35--68 s (filled circles) in the
  detuning energy ranges of (a) $E_d=1$--15 eV and (b) $E_d=15$--40
  eV.}
\label{dr_rate_coeff02}
\end{figure}

At high collision energies, the variation of the rate coefficient
$\tilde{\alpha}_{\rm DR}(E_d)$ directly reflects the energy dependence
of the DR cross section.  The electron energy spread ($\sim$0.23 eV at
10 eV and $\sim$0.34 eV at 20 eV) is small compared to the observed
structures, so that the DR cross section $\tilde{\sigma}_{\rm DR}(E)$
can be obtained directly from the rate coefficient using the relation
$\tilde{\alpha}_{\rm DR}(E_d)\sim (2E_d/m)^{1/2}\tilde{\sigma}_{\rm
  DR}(E_d)$.  The high-energy structure of the measured rate
coefficient (Fig.\ \ref{dr_rate_coeff02}) shows much less variation
with the storage time, indicating that the sizes of the state-specific
cross sections are similar to each other.  Between 3 and 12 eV, a
broad peak structure is observed that, as shown in Fig.\ 
\ref{dr_rate_coeff02}(a), slightly shifts and narrows with time, again
reflecting the evolution of rovibrational states in the beam during
storage.  The position and the width of this structure match well
the energies of vertical transitions from the ionic vibrational ground
state to dissociating neutral Rydberg states which converge to the
repulsive $A^{2}\Sigma_u^+$ ionic state shown Fig.\ \ref{he_curves}.
The DR cross section arising from this reaction channel is found to be
$(1.7\pm0.2)\times10^{-16}$ cm$^{2}$ at the peak near 7.3 eV.

At higher energies ($>$15 eV) the measured rate coefficient displays a
rich structure with a small narrow peak (FWHM $\sim$2 eV) at $\sim$19
eV, a larger and broader peak at $\sim$22.5 eV, a plateau-like
behavior at 25--27 eV, and a small modulation at 31 eV before the rate
declines.  
In the energy region of these observed structures, vertical
transitions from the ionic vibrational ground state can
reach dissociating neutral Rydberg states attached to highly excited
states of the He$_2^+$ ion; the ionic curves correlated
to the He$^+(1s)$ + He$(1s2l)$ atomic limits are shown in 
Fig.\ \ref{he_curves-high}.  
For instance, the energetic limit for dissociation 
into He$(1s2s)$ + He$(1snl)$ ($n\ge2$) is 17.4 eV, 
and the peak at $\sim19$ eV could well represent dissociation 
to these final states following an initial vertical electron 
capture process. 
Similarily, the energetic limit for dissociation into 
He$(1s3l)$ + He$(1snl)$ ($n\ge3$) is  23.6 eV, for 
He$(1s4l)$ + He$(1snl)$ ($n\ge4$) 25.0 eV, while the
energetic limit for He$^+(1s)$ + He$^+(1s)$ is found 
at 26.9 eV. 
The very detailed structure, in particular the
narrow peak at $\sim$19 eV, may indicate the importance 
of individual core excited resonances as in the case 
of CD$^+$ \cite{forck1994}.

\subsection{Energetic structure of DE and electron-impact
  vibrational excitation}
\label{sec:res_de_exc}

\begin{figure}[htbp]
\centering
\includegraphics[width=3.0in]{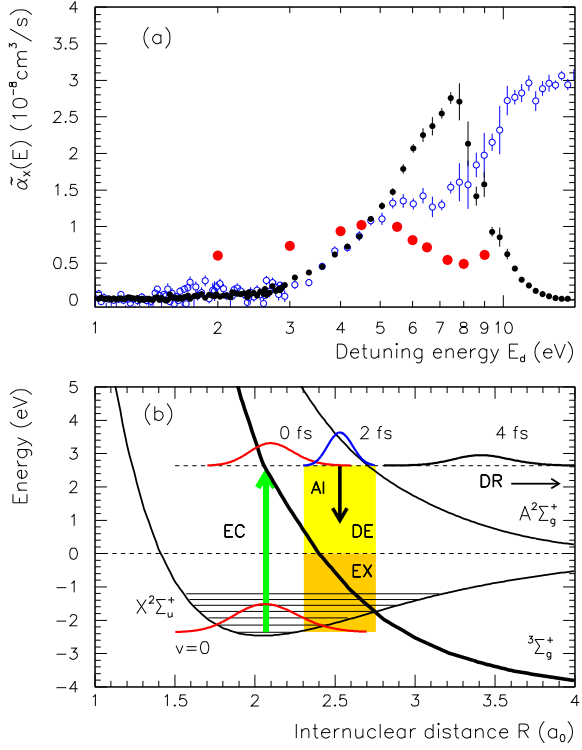}
\caption{
  (a) Measured rate coefficients for DR (black filled circles) and DE
  (open circles) at $t=35$--68 s together with the vibrational
  excitation profile as obtained from the pump-probe experiments
  (light filled circles).  The absolute scale of the excitation
  profile is arbitrarily chosen to compare its shape to the rate
  coefficients for DR and DE. 
  (b) Schematic illustration (inspired by Ref. \cite{orel1996})
  of the DR, DE and vibrational excitation 
  (EX) processes following an initial electron capture (EC) at 5 eV from 
  the vibrational ground state ($v=0$) into the ${^3}\Sigma_g^+$ 
  dissociative Ryberg state. 
  Immediately after electron capture a nuclear wave packet 
  (marked 0 fs) is formed in the ${^3}\Sigma_g^+$ potential, 
  where it is repelled towards larger internuclear 
  distances as illustrated by the wave packets at 2 fs and 
  4 fs; a process that finally leads to DR. During repulsion, 
  the wave packet looses intensity since the molecule 
  can autoionize (AI) to the ionic ground state, leading to either DE or EX. 
  The gray shaded areas show the Frank-Condon region for electron
  reemission for the wave packet at 2 fs.}
\label{ex_de_dr}
\end{figure}

In the present experiment, all three types of
electron-induced reactions discussed in Sec.\ \ref{sec:intro}
were studied as a function of the collision energy. For
the energy range of vertical transitions from the ionic ground state
to neutral dissociating Rydberg states attached to the lowest excited
state of He$_2^+$ ($\sim$3--10 eV) the measured reaction rate
coefficients for DE and DR are compared in Fig.\ \ref{ex_de_dr}(a). 
The energy dependence of the signal for electron-impact vibrational
excitation to states $v\geq3$, as obtained from the pump-probe
measurements, is also included in Fig.\ \ref{ex_de_dr}(a). 
The absolute scale of the excitation profile is arbitrarily chosen in
order to compare its shape to the rate coefficients for DR and DE.

Considering only the ``direct'' DE caused by simple electron impact
excitation to the repulsive curve of He$_2^+$, the DE cross section
should rise only at $\gtrsim$5 eV (cf.\ Fig.\ \ref{he_curves}).  Here,
we observe that the DE rate coefficient rises already at much lower
energies, close to the dissociation threshold of $\sim$2.4 eV, and
indeed follows the energy dependence of the DR rate coefficient up to
$\sim$5 eV.  Above this energy, the DE rate ceases to rise until the
onset of the ``direct'' DE at $\sim$7 eV, which leads to a strong
increase towards higher energies, while the DR rate drops.
The DE signal below $\sim$7 eV clearly shows the influence of
intermediate neutral Rydberg states on this process.
The vibrational excitation signal also occurs  below the
dissociation threshold and it extends far above the dissociation
energy with a peak at $\sim$5 eV.  The peak position matches the
vertical transition energy from the ionic ground state to the first
neutral dissociative curve $^3\Sigma_g^+$ [see Fig.\ \ref{ex_de_dr}(b)],
but also the higher lying neutral states below the $A^{2}\Sigma_u^+$
ionic state appear to contribute. A similar excitation profile was
previously reported for HD$^+$ \cite{lange_phd,zajfman2003}.
The results underline the importance of the electronically doubly
excited states in all, DR, DE, and electron-impact excitation of
He$_2^+$ \cite{nakashima1986}. 

A mechanism of resonance enhanced DE and vibrational excitation 
was described by Orel and Kulander \cite{orel1996} for HeH$^+$.
A similar mechanism is illustrated in Fig.\ \ref{ex_de_dr}(b) for 
He$_2^+$ involving the first ${^3}\Sigma_g^+$ Ryberg state reached 
vertically at 5 eV above the vibrational ground state.
Following
resonant capture of the incident electron, the resonance state autoionizes
with a considerable fraction of the 
excitation energy being given to the nuclear motion, leading either 
into the vibrational continuum (DE) or into a bound vibrational
state (EX) of the He$_2^+$ ion in its electronic ground state.
DR results if the resonance evolves without autoionization.

\section{Discussion}

\subsection{${^3}$He${^4}$He$^+$ DR}
\label{sec:discussion_dr}

\subsubsection{DR rate coefficient}


Our general observations regarding the DR rate coefficient of
$^3$He$^4$He$^+$ and its dependence on the ion storage time are in
overall agreement with those of the only earlier storage ring
experiment at ASTRID \cite{urbain1999b}, which also showed a strong
variation of the low-energy DR rate during the vibrational relaxation
and indicated a small rate coefficient for the vibrational ground
state.  Very recently, absolute rates from this experiment were
presented \cite{urbain2004}.  For low-energy DR, a thermal rate
coefficient of $(6\pm3)\times 10^{-10}$ cm$^3$\,s$^{-1}$ for 300 K was
extracted from the data at $\gtrsim$12 s of storage time, including
all possible contributions from excited ro-vibrational states of the
stored ions.  Within the statistical errors, the fragment imaging 
distributions from this experiment, also reported recently \cite{urbain2004}, 
do not show any contribution of the $n=2$ final atomic states, 
unlike the present data (cf.\ Sec.\ 
\ref{initial_final}).  Hence, the extraction of a rate coefficient for
$v=0$ ions is not possible there; moreover, this may indicate a higher
internal excitation of the stored ions than in our case.  Converting
our zero-energy, rovibrationally averaged DR rate coefficient
$\tilde{\alpha}_{\rm DR}(0)=2.7(4)\times 10^{-9}$ cm$^3$\,s$^{-1}$ to
a 300-K thermal value through the division by 2.2 (see Sec.\ 
\ref{sec:absrate}), we obtain---as a quantity equivalent to the
rovibrationally averaged ASTRID result---a value of $1.2\times
10^{-9}$ cm$^3$\,s$^{-1}$, significantly higher in spite of probably a
lower internal excitation of the ions.  
Regarding high-energy DR, we
find a significantly larger cross section than ASTRID on the peak near
7 eV [$1.7(2)\times 10^{-16}$ cm$^2$ (see Sec.\ 
\ref{sec:high-e-structure}) vs.\ $0.6(3)\times 10^{-16}$ cm$^2$ at
ASTRID]; in addition, different values are found for the position of
the maximum (7.3 eV vs.\ 6.6 eV at ASTRID).  While some of these
discrepancies may be explained by the different ion source and storage
conditions in the two experiments, the reasons for the apparent
disagreement in the absolute cross section, particularly obvious for
the high-energy peak, are presently unknown.
 
For the vibrational ground state, our thermal DR rate coefficient of
$\tilde{\alpha}^{v=0}_{{\rm 300\,K,~DR}} = (3.3 \pm 0.9) \times
10^{-10}$ cm$^3$\,s$^{-1}$ is consistent with the limit of
$<${}$5\times10^{-10}$ cm$^3$\,s$^{-1}$ obtained by Deloche {\em et
  al.} \cite{deloche1976}.  For excited vibrational states, the result
of $\tilde{\alpha}^{v\geq3}_{\rm DR}(0) \gtrsim 2\times10^{-7}$
cm$^{3}$\,s$^{-1}$ (Sec.\ \ref{sec:absrate}), corresponding to thermal
rate coefficients of $\gtrsim 1\times10^{-7}$ cm$^{3}$\,s$^{-1}$,
clearly exceeds the limit of Ref.\ \cite{deloche1976}.  However, the
observed strong increase of the DR rate coefficient for excited
vibrational states is in accord with the measurement by Ivanov {\it et
  al.}  \cite{ivanov1989} and with the calculations by Carata {\it et
  al.}  \cite{carata1999}.

Our low-energy thermal rate coefficient for the vibrational ground
state is about a factor of 6 larger than the theoretical result
($6.1\times10^{-11}$ cm$^3$\,s$^{-1}$) of Carata {\it et al.}
\cite{carata1999} which, although given for $J=0$, differs only little
from that for a rotational temperature of 300\,K (see Fig.\ 14 of
Ref.\ \cite{carata1999}) which should describe the present
experimental conditions more adequately.  Carata {\it et al.}
emphasize \cite{carata1999} that non-adiabatic couplings to singly
excited neutral states are not included in their calculation, so that
the larger rate coefficient observed here may indicate the presence of
such couplings (however, possible branching to ground-state He atoms
\cite{carata1999} can be excluded experimentally as discussed in the
following Section).

Very recently, Royal and Orel \cite{royal2004} have obtained results
for the high-energy DR of ${^3}$He${^4}$He$^+$ ($\sim$2--13 eV) using
time dependent wave packet calculations.  Their peak cross section of
$0.75\times 10^{-17}$ cm$^2$ is significantly smaller than our result;
the peak position is in good agreement, while the calculated peak
shape does not reproduce the asymmetry found in our data for $t>35$ s
(see Fig.\ \ref{dr_rate_coeff02}, and Fig.\ 3 of Ref.\ 
\cite{royal2004}).  The latter observation may indicate the presence
of mechanisms or dissociating curves not included in the calculation.

\subsubsection{DR branching ratios}
\label{sec:disc_branching}

To the best of our knowledge, previous experimental results
regarding the final-state branching ratios for the low-energy DR of
$v=0$ He$_2^+$ ions suitable for a comparison with our results are not
available.  Carata {\it et al.}  \cite{carata1999} in their
calculations found the $^{3} \Sigma_g^+$ curve to be the dominant
dissociation route for the vibrational ground state.  Although this
curve diabatically correlates to the $1s2s\,^3S$ final atomic state,
we find only $\sim$3.7\% of the DR events from the vibrational ground
state in this channel (see Table \ref{table1}).  The experimentally
strongest channels ($^3P$ and $^1S$) correspond to dissociating curves
yielding only minor contributions to the calculated DR rate
\cite{carata1999}.  In the context of these calculations this would
indicate the presence of substantial population exchange between
molecular states in the dissociation process.  Such effects, which
clearly need further study, could be caused by the interference of the
three dissociation routes ($^3\Sigma_g$, $^3 \Pi_u$, and $^1\Sigma_g$)
through $\Sigma$-$\Pi$ coupling \cite{cohen1976} or by
non-adiabatic couplings \cite{guberman1994,sarpal1994}.  An example of
a similar population exchange can be found in the DR of O$_2^+$
\cite{kella1997,guberman1997}.

The inclusion of non-adiabatic couplings, suggested by the difference
between the experiment and the MQDT result for the $v=0$ DR rate, as
well as by the possible re-population effects considered just above,
was suspected \cite{carata1999} to give more flux of dissociation to
energetically lower asymptotic channels, notably the ground state
product channel He$(1s^2)$ + He$(1s^2)$.  From our DR fragment imaging
results, we can with high sensitivity exclude the presence of this
channel, which for our experimental parameters would have a maximum
interparticle separation of 37.2 mm and thus be easily recognizable on
the 80-mm diameter fragment imaging detector.  In fact, also in the
case of HeH$^+$ DR, characterized by strong non-adiabatic coupling,
ground state products could be excluded with high experimental
sensitivity \cite{semaniak1996}.

The small contribution from higher vibrational states in Region 2 of
the fragment imaging distributions (see Sec.\ \ref{initial_final}) is
remarkable.  Consequently, the strong DR from excited vibrational
levels ($v \ge 3$) proceeds mostly to final states with an excited
He($nl$) atom with $n\ge3$.  Similar strong preferences for DR into
the highest energetically open final channels were observed previously
for the DR of HeH$^+$ \cite{semaniak1996} and LiH$^+$
\cite{krohn2001}.

\subsection{Rovibrational excitation and de-excitation mechanisms and rates}    
\label{sec:resultsC}
\label{sec:discussion_rovib}

\subsubsection{Vibrational excitation of stored $^3$He$^4$He$^+$ ions}
\label{sec:modelcalc_vib}

\begin{figure}[htbp]
\centering
\includegraphics[width=3.0in]{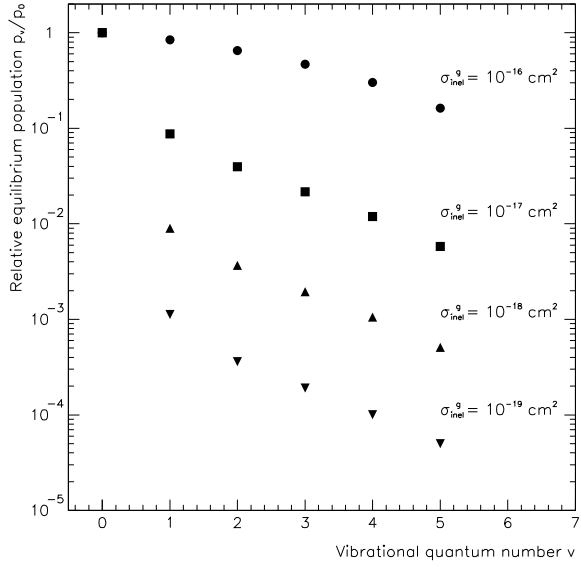}
\caption{
  Calculated vibrational equilbrium populations $p_v$, relative to the
  population $p_0$ in $v=0$, obtained when vibrational excitation
  through collisions with the residual gas [Eq.\ (\ref{gas_model})] is
  included in the model of Eqs.\ (\ref{modelevolution}) for four
  values of the excitation cross section $\sigma^g_{\rm inel}$ as
  indicated.  Six vibrational levels ($v=0$-5) and thirty rotational
  levels ($J=0$--29) on each vibrational level are modeled; the
  residual gas density is set to $n_g = 1.3\times 10^{6}$ cm$^{-3}$
  and the temperature of the radiation field to 300 K.}
\label{gas_asymptote}
\end{figure}

Using the coupled set of Eqs.\ (\ref{modelevolution}) for the
rovibrational populations in the stored ion beam, we have modeled the
effect of inelastic collisions of the stored ions with residual gas
molecules acting in addition to the blackbody-induced radiative
transitions already discussed in Sec.\ \ref{sec:Ion_beam_evolution}.
Lacking measurements or calculations 
suitable for a realisic modeling of MeV collisions between molecular
ions and H$_2$ (the main component of the residual gas),  
we strongly simplify the case, using a single fixed cross section to
describe the probability that the state $v,J$ of a molecular ion is
changed in a residual gas collision to a different state $v',J$ with a
certain $v'\neq v$.  The same partial cross section $\sigma^g_{\rm
  inel}$ is used for all possible final $v'$ within a model containing
six vibrational levels ($v=0$--5) and thirty rotational levels
($J=0$--29).  Since the relevant collisions, leaving the molecular ion
intact in spite of the high collision energy, are likely to proceed at
large distance with a small exchange of momentum, we assume that the
rotational quantum number is preserved.  We then obtain rate constants
$k^g_{\rm inel}$ according to Eq.\ (\ref{eq:kg_general}) with $n_g=
1.3 \times 10^{6}$ cm$^{-3}$ and include the term
\begin{equation}
{} + \sum_{v'} ( k^g_{\rm inel} N_{v'J} - k^g_{\rm inel}  N_{vJ} )
\label{gas_model}
\end{equation}
on each of the right-hand sides of Eqs.\ (\ref{modelevolution}), thus
describing the collisional excitation and de-excitation of vibrational
levels in residual gas collisions.  As expected (see Sec.\ 
\ref{sec:Ion_beam_evolution}) the equilibrium under the influence of
residual gas collisions was always reached faster than under the
action of radiative transitions alone (Fig.\ \ref{radiative1}).  The
equilibrium populations obtained in competition with radiative
relaxation for four different values of $\sigma^g_{\rm inel}$ are
shown in Fig.\ \ref{gas_asymptote}.  The excited-state equilibrium
populations on the order of 1\% indicated by the experimental results
are found to correspond to partial inelastic collision cross sections
of the order of a few 10$^{-18}$ cm$^2$.

From the ion beam decay rate $k_{\rm DE}^g$ (see Sec.\ 
\ref{sec:drtime}) the partial cross section for each of the two DE
reaction channels in a residual gas collision [Eq.\ (\ref{eq:de_res})]
is determined to be $2.7\times10^{-17}$ cm$^2$.  On the other hand,
the cross section for dissociative charge exchange [Eq.\ 
(\ref{eq:dc_res})] is found to be $4\times10^{-20}$ cm$^2$.  
The estimated vibrational excitation cross sections lie between 
both values.  The
reaction mechanism(s) that can lead to vibrational excitation of the
stored ions after collisions with a residual gas molecule (H$_2$) at
MeV energies is not obvious.  It may proceed as a direct vibrational
excitation reaction or it may proceed as a relaxation process after an
initial charge exchange collision.  Clearly, to understand better the
importance of the vibrational excitation in storage ring measurements,
there is a need for theoretical work on such processes both in view of
obtaining absolute cross sections and understanding the underlying
mechanisms.

The result shows that for molecular ions that are not or only weakly
radiatively active  (unlike $^3$He$^4$He$^+$), an ion beam stored in a 
storage ring may stabilize
with a significant vibrational excitation, even though the molecules
may have been produced vibrationally cold in the ion source.  However,
in the well studied case of stored H$_2^+$ ions, electron induced
vibrational de-excitation turned out to be a strong process
\cite{krohn2000} and found to dominate the stabilized level of
vibrational excitation.

\subsubsection{Rotational excitation and cooling}
\label{sec:modelcalc_rot}

The interaction with electrons at zero detuning energy ($E_d=0$) was
shown in Sec.\ \ref{sec:total-DR-rate} to have a strong effect on the
DR rate coefficient $\tilde{\alpha}_{\rm DR}(0)$, as represented by
the time dependence of the normalized DR rate $\rdr/\rdei$. 
While the evolution $\lesssim10$ s was attributed mainly to vibrational 
stabilization, the time scale of the observed variations beyond 10 s 
was assigned to changes in the relative rotational level populations 
within the vibrational ground state.  
Here we address the time dependence of the normalized DR rate 
$\rdr/\rdei$ beyond 10 s by exploring possible mechanisms changing 
the rotational populations in the stored ion beam in a similar way 
as in the previous subsection, extending the set of 
Eqs.\ (\ref{modelevolution}) and following the 
resulting time evolution in a model calculation.
A complete modeling of the normalized DR rate $\rdr/\rdei$ is excluded 
considering that the detailed rate coefficients $\alpha_{\rm DR}^{vJ}$ 
(possibly with strong resonances for certain $J$'s) are unknown.
With the aim of illustration, we focus mainly on the time dependence of
the rotational temperature for an ensemble of ions, since the time scale 
of changes in the normalized DR rate $\rdr/\rdei$ still qualitatively 
reflects the time scale of changes in the rotational temperature, 
despite a complicated $J$-dependence of $\alpha_{\rm DR}^{vJ}$.

\begin{figure}[htbp]
\centering
\includegraphics[width=3.0in]{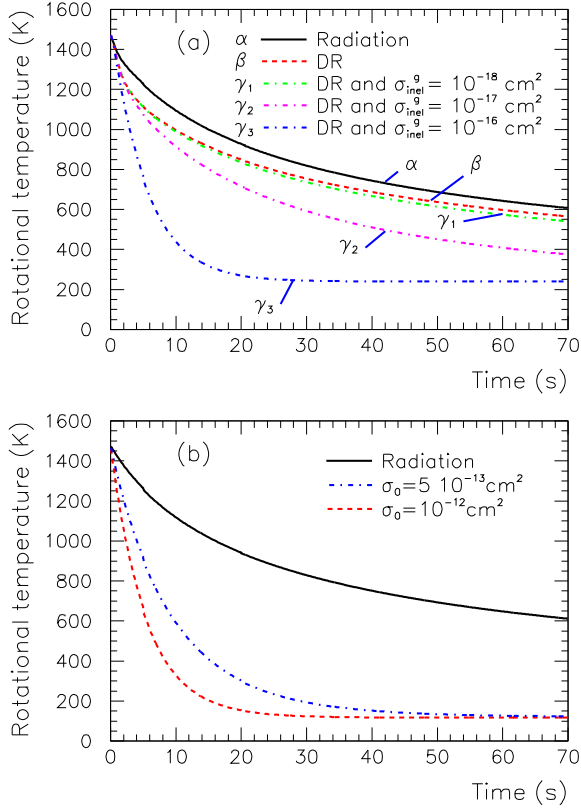}
\caption{
  Calculated effects of electron-ion collisions and ion-residual gas 
  collisions on the rotational
  thermalization in the $X{^2}\Sigma_u^+$ electronic ground state of
  $^3$He$^4$He$^+$ from an initial temperature of 1500 K.  (a)
  Influence of ($\alpha$) radiative transitions (solid), ($\beta$)
  radiation and DR depletion [Eqs.\ (\ref{depletion}),
  (\ref{depletion_model}); dashed] and ($\gamma_{1\!-\!3}$) radiation,
  DR depletion, and vibrational excitation in the residual gas [Eq.\ 
  (\ref{gas_model})] with different excitation cross sections
  $\sigma^g_{\rm inel}$ (dotted).  Initially, the six lowest
  vibrational levels were populated equally, with a rotational
  temperature of 1500 K imposed on each level.  The electron density
  was set to $n_e= 1.2 \times 10^{7}$ cm$^{-3}$ and the residual gas
  density to $n_g = 1.3\times 10^{6}$ cm$^{-3}$.  (b) Influence of
  radiative transitions (solid) and of radiation plus rotationally
  inelastic collisions in the velocity-matched electron beam [Eqs.\ 
  (\ref{rot_cross1})--(\ref{rot_model})].  The cross section constant
  was set to $\sigma_0=5\times10^{-13}$ cm$^2$ (dash-dotted) and
  $10^{-12}$ cm$^2$ (dashed).  Initially, only the vibrational ground
  state was assumed to be populated with a rotational temperature of
  1500 K.  The experimental electron temperatures were used to derive
  the rate coefficients of the inelastic electron collisions; $n_e$ as
  in (a).}
\label{electron}
\end{figure}

As discussed in Sec.\ \ref{sec:total-DR-rate}, a rotational dependence of
the DR rate coefficient might lead to selective depletion of excited
rotational levels, provided the involved rate coefficients are large
enough.  To schematically investigate this effect, we parameterize the
DR rate coefficient as
\begin{equation}
\alpha_{\rm DR}^{vJ}(0) = a_v (1 + bJ),
\label{depletion}
\end{equation}
where the parameter $a_v$ represents the vibrational and $b$ the
rotational dependence.  The $v$- and $J$-dependent DR rates are
included by adding the terms
\begin{equation}
{} - k^{vJ}_{\rm DR}(0) N_{vJ}
\label{depletion_model}
\end{equation}
to the right-hand sides of Eqs.\ (\ref{modelevolution}).  The increase
of the DR rate with the vibrational level was included by chosing
$a_0=5\times10^{-10}$ cm$^3$\,s$^{-1}$ and letting $a_v$ increase by a
factor of 10 for each consecutive excited state up to $v=3$, and then
setting $a_v=5\times10^{-7}$ cm$^3$\,s$^{-1}$ for $v\geq3$.  A strong
$J$-dependent increase of $\alpha_{\rm DR}^{vJ}(0)$ was assumed with
$b=1$.  As before, a model with six vibrational levels ($v=0$--5) was
considered.  In comparison to radiative relaxation alone [Fig.\ 
\ref{electron}(a), curve $\alpha$], the inclusion of DR depletion
(curve $\beta$) has a small effect as long as the vibrational
excitation in residual gas collision is neglected (depletion is only
efficient at short times, when excited vibrational levels are
still significantly populated).  If the population of excited
vibrational levels by residual gas collisions is included, adding the
terms of Eq.\ (\ref{gas_model}), the depletion effect tends to become
stronger; however, a time scale of the cooling effect comparable to
that observed in Fig.\ \ref{cooling_fig1} is reached only for very
large values of $\sigma^g_{\rm inel}$, for which excited vibrational
states would be about as strongly populated as the $v=0$ ground state.
This situation is not found in the experiment and hence, as already
stated in Sec.\ \ref{sec:total-DR-rate}, depletion [even with the
strong effect implied by choosing $b=1$ in Eq.\ (\ref{depletion})]
cannot lead to significant changes of the rotational temperature on
the time scales of the electron-induced cooling effect of Fig.\ 
\ref{cooling_fig1}.

To estimate the effect of rotationally inelastic electron
collisions [Eq.\ (\ref{eq:general_ex}) with $v=v'$] we parameterize
the rotational excitation cross section, inspired by the formulation
given by Rabadan {\it et al.}\ \cite{rabadan1998a}, as
\begin{equation}
\sigma_{\rm EX}^{J \rightarrow J'}(E) =
  \sigma_0 \sqrt{\frac{E-E_{J'}^{J}}{E}} \frac{E_{J'}^{J}}{E},
\label{rot_cross1}
\end{equation}
where the overall size of the cross section is determined by the
parameter $\sigma_0$.  The corresponding de-excitation cross section
is derived from the principle of microscopic reversibility, and
written as
\begin{equation}
\sigma_{\rm DEX}^{J' \rightarrow J}(E') =
 \sigma_{\rm EX}^{J \rightarrow J'}(E'+E_{J'}^{J}) 
  \sqrt{\frac{E'+E_{J'}^{J}}{E'}} \frac{2J+1}{2J'+1}.
\label{rot_cross2}
\end{equation}
The rotationally inelastic electron collisions are then included in
the coupled set of Eqs.\ (\ref{modelevolution}) by adding the following
terms to its right hand side
\begin{eqnarray}
& + &\sum_{(vJ')>(vJ)} \left( k_{\rm DEX}^{J' \rightarrow J}(0)N_{vJ'}  -
                              k_{\rm EX}^{J \rightarrow J'}(0) N_{vJ}  \right)
\nonumber\\
& + &\sum_{(vJ')<(vJ)} \left( k_{\rm EX}^{J' \rightarrow J}(0) N_{vJ'} -
                              k_{\rm DEX}^{J \rightarrow J'}(0)N_{vJ}  \right),
\label{rot_model}
\end{eqnarray}
where $k_{\rm EX}^{J \rightarrow J'}(E_d)$ and $k_{\rm DEX}^{J
  \rightarrow J'}(E_d)$ are defined according to Eq.\ 
(\ref{eq:kx_general}).  Figure \ref{electron}(b) illustrates the
effect of the electron induced rotational transitions for two
different values of $\sigma_0$ in comparison to the case where only
radiative transitions affect the rotational population.  For
simplicity, the electron induced transitions included in the model
were restricted to $\Delta J= \pm 1$. 

\begin{figure}[htbp]
\centering
\includegraphics[width=3.0in]{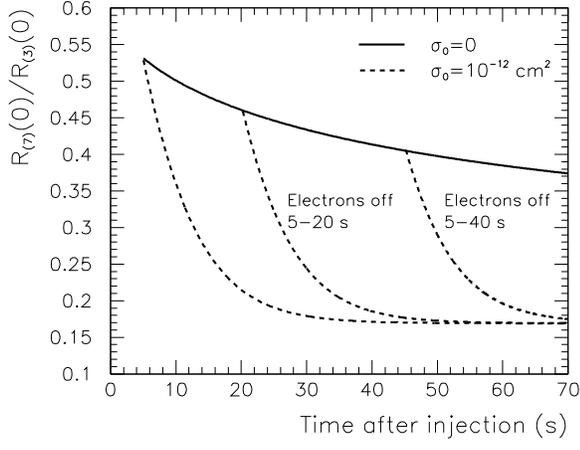}
\caption{
  Model calculation of the observable ratio $\rdr/\rdei$ 
  including the effect of radiation, DR depletion, and 
  rotationally inelastic collisions.  
  Only the vibrational ground state v = 0 was populated and it was 
  assumed that a rotational temperature of 1500 K was 
  reached after 5 s of ion storage.
  The electron density was $n_e= 1.2 \times 10^{7}$ cm$^{-3}$. 
  $\rdr$ was obtained using the model rate coefficient 
  of Eq. \ (\ref{depletion}) with $a_0=5\times10^{-10}$ cm${^2}$s$^{-1}$ 
  and $b=1$. 
  $\rdei$ rate was calculated as 
  $(f_g k_{\rm DE}^g + \tilde{k}_{\rm DE}(0))\times N_i$, with 
  $f_g=0.044$, $k_{\rm DE}^g=$ 0.0506 s$^{-1}$, 
  and  $\tilde{k}_{\rm DE}(0)=8.73\times10^{-4}$ s$^{-1}$.
  The solid curve shows the ratio $\rdr/\rdei$ when the  
  rotationally inelastic collisions are neglected ($\sigma_0=0$).	
  The dashed curves show the ratio $\rdr/\rdei$ computed with a 
  strong electron induced rotational de-excitation 
  ($\sigma_0=10^{-12}$ cm$^2$) 
  with the electron beam being either continously on or switched off 
  for time intervals of  5-20 s and 5-40 s as marked.}
\label{model_rates}
\end{figure}

This calculation indicates that significant changes of the rotational
temperature on the time scales of the observed cooling effect in Fig.\ 
\ref{cooling_fig1} can be obtained by electron induced rotational
transitions for $\sigma_0$ of the order of $10^{-12}$ cm$^2$; the
collisional cooling in fact leads down to the electron temperature
($kT_{\perp}=10$ meV, $T_{\perp} = 116$ K) which is included through
the averaging of the excitation and de-excitation cross sections
according to Eq.\ (\ref{eq:kx_general}).  The required value of the
cross section constant $\sigma_0$ (amounting to $\sim$10$^4$ \AA$^2$)
is not unrealistic \cite{rabadan1998b} and slightly smaller values may
in fact be sufficient if transitions other than $\Delta J = \pm 1$
\cite{rabadan1998a} are considered in addition.
Fig.\ \ref{model_rates} displays calculated ratios $\rdr/\rdei$ as a function 
of time both when the rotational de-excitations are negligible (soild line) 
and when they have a strong effect on the rotational temperature (dashed lines).
Even with the simplifying assumptions made here, the model calculation
in Fig.\ \ref{model_rates} reproduces almost quantitatively the experimental 
data of Fig.\ \ref{cooling_fig1} when a strong rotational de-excitation is 
included.


Summarizing, the model calculations---in combination with the finding
of only small populations on the per-cent level in excited vibrational
states---suggest that the observed temporal evolution of the
low-energy DR signal of stored $^3$He$^4$He$^+$ ions is due to
rotational cooling through inelastic electron collisions, and not due
to the combined action of DR depletion and vibrational excitation.  

Rotational excitation and de-excitation in electron collisions have
already been considered theoretically for several molecular ions (see,
e.g., Ref.\ \cite{rabadan1998a}).  It must be expected that such
reactions play a significant role for other storage ring measurements
on DR too.  In a study of D$_2$H$^+$ \cite{lammich2003} a similar time
dependence of the low-energy DR rate was observed, in that case with a
considerably large DR rate coefficient, so that the observed changes
of rotational populations were attributed mainly to selective ion
depletion by DR from the ion beam, implying a linear dependence of the
rate coefficient on the rotational temperature with slope $0.94 \times
10^{-9}$ cm$^3$\,s$^{-1}$K$^{-1}$.  However, additional de-excitation
of D$_2$H$^+$ ions by rotationally inelastic electron collisions could
not be excluded, as also in this case the actual dependence of the DR
rate coefficient on the rotational level is still unknown.


\section{Conclusion}

The present merged-beams DR measurements with $^3$He$^4$He$^+$ ions
clearly confirm the predicted strong dependence of the DR cross
section for He$_2^+$ on the initial vibrational level.  DR
observations at low collision energies ($\lesssim$100 meV) have turned
out to be highly sensitive to small populations in vibrationally
excited states.  Although $^3$He$^4$He$^+$ is radiatively active,
which should result in negligible populations in vibrationally excited
states after already a few seconds of storage, evidence for
stationary, non-thermal populations on vibrational excited levels with
fractions of 0.1 to 1\% was found by analyzing the DR fragment imaging
distributions.  Pump-probe-type measurements involving switching or
variations of the energy of the electron beam allowed the vibrational
excitation by energetic electron-ion collisions to be distinguished
from that by ion collisions with the residual gas, finding the latter
to dominate in the normal storage conditions.

On the longer time scales typical for the evolution of rotational
level populations, the presence of the velocity-matched electron beam
is seen to cause significant changes of the observed (state-averaged)
zero-energy DR rate coefficient, which are attributed to rotational
cooling of the $^3$He$^4$He$^+$ ions in connection with a
dependendence of the DR cross section on their initial rotational
state.  The cooling effect could be unambiguously attributed to
rotational de-excitation in inelastic low-energy electron-ion
collisions.  Selective depletion by a rotationally sensitive loss
process can be excluded since the only process which could provide for
such a sensitivity is low-energy DR, and the corresponding rate
coefficient is observed to be two orders of magnitude too low to
account for sufficient rotational population changes.

With the developed understanding of the beam dynamics and the
contributions of different reaction channels to the DR fragment
imaging distribution, an absolute DR rate coefficient of $(7.3 \pm
2.1) \times 10^{-10}$ cm$^{3}$\,s$^{-1}$ could be determined for $v=0$
$^3$He$^4$He$^+$ ions, referring to an electron thermal energy near 10
meV and to a rotational temperature likely to be near 300 K or below.
The result is somewhat larger than predicted in recent theoretical
calculations \cite{carata1999} and consistent with earlier
experimental limits \cite{deloche1976}.  Also branching ratios for the
four accessible final atomic states could be determined (Table
\ref{table1}); they significantly differ from those expected
\cite{carata1999} on the basis of the relative importance of the
various dissociative potential curves and their diabatic correlations
to final atomic levels.

In an energy range up to 40 eV, the energy dependences of the cross
sections for DR, DE and electron-impact vibrational excitation were
measured.  The DR cross section shows the expected broad peak at
$\sim$7.3 eV and interesting structures, one of them narrow, at higher
energies, which appear to carry detailed information about higher
lying doubly excited states of He$_2$ and call for further
investigations.  The measurements also reveal the competition between
different stabilization pathways after the resonant capture of an
electron on He$_2^+$ at energies of several eV.

In a broader view, the results presented in this paper illustrate some
of the challenges faced by storage ring experiments setting out to
measure very small DR cross sections or to obtain sensitivity on
initial rotational states in such measurements.  At the present stage,
in particular the rotational excitation can only be extracted by
indirect conclusions bringing together several experimental findings.  
In the future, more direct diagnostic techniques to study the
rotational population of the stored and electron cooled ion beams,
similar to the laser technique demonstrated earlier for a specific
case \cite{hechtfischer1998}, are highly desirable.  With such
diagnostic techniques established, the strong rotational excitation
and de-excitation that appears to be possible by low-energy collisions
in the merged electron beam could also be used to actively manipulate
the rotational populations in a stored ion beam.



\begin{acknowledgments}

This work has been funded by the German Israel Foundation 
for Scientific Research (GIF) under Contract No. I-707-55.7/2001 
and by the European Community within the Research Training Network 
``Electron Transfer Reactions''.
HBP acknowledges support from the European Community program IHP
through a Marie Curie fellowship under contract
No.\ HPMF-CT-2002-01833.

\end{acknowledgments}



\clearpage

\newpage
\begin{table}
\caption{\label{tableB} 
Signal contributions (normalized to the integral of the observed
distribution) obtained from a least-squares fit 
to the DR fragment imaging distribution at $E_d=0$, listed in the 
sequence of increasing kinetic energy release.}
\begin{ruledtabular}
\begin{tabular}{c@{\hspace{4mm}}c@{\hspace{-5mm}}d}
 Initial state & Final channel & \multicolumn{1}{r}{Contribution} \\
        \hline
\raisebox{0mm}[4mm][0mm]{}%
$^{3}$He$^{4}$He$^+$($v=0$) & He($1s^2 \, ^{1}S$) + He($1s2p\,^{1}P$) & 0.0154(16)  \\
$^{3}$He$^{4}$He$^+$($v=0$) & He($1s^2 \, ^{1}S$) + He($1s2p\,^{3}P$) & 0.311(28) \\
$^{3}$He$^{4}$He$^+$($v=0$) & He($1s^2 \, ^{1}S$) + He($1s2s\,^{1}S$) & 0.198(21) \\
$^{3}$He$^{4}$He$^+$($v=3$) & He($1s^2 \, ^{1}S$) + He($1s2p\,^{3}P$) & 0.041(14) \\
$^{3}$He$^{4}$He$^+$($v=0$) & He($1s^2 \, ^{1}S$) + He($1s2s\,^{3}S$) & 0.0196(6) \\
$^{3}$He$^{4}$He$^+$($v=4$) & He($1s^2 \, ^{1}S$) + He($1s2s\,^{3}S$) & 0.025(3) \\
\end{tabular}
\end{ruledtabular}
\end{table}

\begin{table}
\caption{
\label{table1} Measured final state branching ratios following 
low-energy DR from the $v=0$ level of $^{3}$He$^{4}$He$^+$}
\begin{ruledtabular}
\begin{tabular}{cc}
 Final channel & \multicolumn{1}{c}{Branching ratio (\%)} \\ 
       \hline
\raisebox{0mm}[4mm][0mm]{}%
   He($1s^2 \, ^{1}S$) + He($1s2s \, ^{3}S$) & $\,~3.7 \pm 1.2$   \\
   He($1s^2 \, ^{1}S$) + He($1s2s \, ^{1}S$) & $37.4 \pm 4.0$   \\
   He($1s^2 \, ^{1}S$) + He($1s2p \, ^{3}P$) & $58.6 \pm 5.2$   \\
   He($1s^2 \, ^{1}S$) + He($1s2p \, ^{1}P$) & $\,~2.9 \pm 3.0$   \\
\end{tabular}
\end{ruledtabular}
\end{table}

\clearpage

\clearpage

%

\clearpage

\end{document}